\documentclass[%
 reprint,
preprintnumbers,
 amsmath,amssymb,
 aps,
]{revtex4-2}

\usepackage{dcolumn}%
\usepackage{bm}%
\usepackage{amsmath}
\usepackage{amssymb}
\usepackage{amsfonts}
\usepackage{amsthm}
\usepackage{amsbsy}
\usepackage{array}
\usepackage{mathtools}
\usepackage[usenames,dvipsnames]{xcolor}
\usepackage{subcaption}
\usepackage{dsdshorthand}
\usepackage{graphicx}
\usepackage[vcentermath]{youngtab}
\usepackage{slashed}

\usepackage{tikz}
\usetikzlibrary{decorations.pathmorphing}
\usetikzlibrary{positioning,decorations.pathreplacing}
\usetikzlibrary{calc}
\tikzset{snake it/.style={decorate, decoration=snake}}
\usepackage{tikz-feynman}

\newcommand{\BFKL}{\textrm{BFKL}}

\newcommand{\QCD}{\textrm{QCD}}

\usepackage{hyperref}%

\makeatletter
\newcommand{\dotbar}{%
  \mathchoice
    {\@dotbar{0.6pt}{0.4em}{1.2pt}}%
    {\@dotbar{0.6pt}{0.4em}{1.2pt}}%
    {\@dotbar{0.4pt}{0.3em}{0.8pt}}%
    {\@dotbar{0.3pt}{0.15em}{0.6pt}}%
}
\newcommand{\@dotbar}[3]{%
  \mathord{\tikz[baseline=-0.5ex]{
    \draw[line width=#1] (0,0) -- (90:#2);
    \draw[line width=#1] (0,0) -- (270:#2);
    \fill (0,0) circle (#3);
  }}%
}

\newcommand{\tripod}{%
  \mathchoice
    {\@tripod{0.6pt}{0.4em}{1.2pt}}%
    {\@tripod{0.6pt}{0.4em}{1.2pt}}%
    {\@tripod{0.4pt}{0.3em}{0.8pt}}%
    {\@tripod{0.3pt}{0.15em}{0.6pt}}%
}
\newcommand{\@tripod}[3]{%
  \mathord{\tikz[baseline=-0.5ex]{
    \draw[line width=#1] (0,0) -- (90:#2);
    \draw[line width=#1] (0,0) -- (330:#2);
    \draw[line width=#1] (0,0) -- (210:#2);
    \fill (0,0) circle (#3);
  }}%
}
\makeatother

\newcommand{\Hdijet}{\cH_{\mkern1mu \dotbar \mkern1mu}}
\newcommand{\Htrijet}{\cH_{\tripod}}
\newcommand{\HdijetJL}{\cH_{J_L}^{\mkern1mu \dotbar \mkern1mu}}
\newcommand{\HtrijetJL}{\cH_{J_L}^{\tripod}}

\begin{document}

\preprint{MIT–CTP 6028 \quad MITP-26-014}

\title{
Operator structure of power corrections and anomalous scaling in energy correlators
}%

\author{Hao Chen}
\email{hao\_chen@mit.edu}
\affiliation{Center for Theoretical Physics - a Leinweber Institute, Massachusetts Institute of Technology, Cambridge, MA 02139, USA}
\author{Yibei Li}
\email{yibei.li@uni-mainz.de}
\affiliation{Mainz Institute for Theoretical Physics,
Johannes Gutenberg University, Staudingerweg 9, 55128 Mainz, Germany}

\begin{abstract}
Energy correlators offer a clean probe of quantum chromodynamics, serving as an ideal laboratory to rigorously investigate non-perturbative power corrections.
The recent discovery that linear corrections exhibit a universal anomalous scaling points to a deep, underlying theoretical structure.
We uncover the quantum field-theoretic origin of this phenomenon in the energy-energy correlator using light-ray operators. 
Through an explicit loop calculation, we derive the one-loop anomalous dimension, revealing that the dijet operator must be combined with a specific triple-jet component. %
This provides a first-principles framework that connects operator theory with high-precision collider phenomenology.
\end{abstract}

\maketitle

\section{Introduction}

Understanding how the fundamental degrees of freedom in Quantum Chromodynamics (QCD) transition into observable hadrons remains one of the defining challenges in theoretical physics.
Energy correlators~\cite{Basham:1978bw,Basham:1977iq,Basham:1979gh,Basham:1978zq} have emerged as a clean and powerful class of observables to study the real-time dynamics of quantum field theories, building a direct bridge between field-theoretical formalisms and high-energy collider experiments (see the review~\cite{Moult:2025nhu}). They can be formulated as correlation functions of light-ray operators along distinct directions~\cite{Hofman:2008ar,Kravchuk:2018htv},  exhibiting elegant symmetry properties~\cite{Kologlu:2019mfz,Chang:2020qpj,Chen:2020adz,Chen:2022jhb,Chang:2022ryc}. Phenomenologically, by measuring the correlation between energy flux, these correlators provide a nice temporal imaging of different phases of QCD evolution at different angular separations~\cite{Komiske:2022enw}. In the past decade, the first-principle perturbative calculations of energy-energy correlator (EEC) in $e^+e^-$ collision has been pushed to unprecedented precision~\cite{DelDuca:2016csb,Tulipant:2017ybb,Dixon:2018qgp,Dixon:2019uzg,Moult:2018jzp,Duhr:2022yyp}. 

Ultimately, the predictive power is limited by non-perturbative power corrections, especially originated from the complex dynamics of hadronization.
For most non-trivial infrared- and collinear-safe (IRC-safe) observables, the typical leading power correction is linear in terms of the inverse of the center-of-mass energy $Q$, {\it i.e.} scaling as $ \Lambda_{\QCD}/Q$, which has been known for more than thirty years~\cite{Dokshitzer:1995zt,Akhoury:1995sp,Nason:1995np,Dokshitzer:1995qm,Beneke:1997sr,Dokshitzer:1997ew,Dokshitzer:1997iz,Dokshitzer:1998pt,Dasgupta:1998xt,Dasgupta:1999mb,Korchemsky:1999kt,Dokshitzer:1999sh,Belitsky:2001ij}. It becomes an indispensable contribution to achieve precise strong coupling constant ($\alpha_s$) measurement from event shapes, including thrust, C-parameter and heavy jet mass~\cite{Nason:2023asn,Bell:2023dqs,Benitez:2024nav,Nason:2025qbx,Aglietti:2025jdj,Benitez:2025vsp}. However, as summarized in~\cite{ParticleDataGroup:2024cfk},  the estimation of these power corrections remains a subject of intense theoretical debate, driven by unresolved discrepancies in the central $\alpha_s$ values extracted across different methodologies.

Recently, several notable studies have revealed the scaling violation properties associated with the hadronization effects~\cite{Mateu:2012nk,Chen:2024nyc,Chang:2025kgq,Lee:2024esz,Lee:2025okn,Guo:2025zwb,Kang:2025zto,Herrmann:2025fqy,Farren-Colloty:2025amh}. In particular, EEC demonstrates its simplicity in the collinear limit, in which power correction and anomalous scaling can be understood using both light-ray operator product expansion~\cite{Chen:2024nyc, Chang:2025kgq} and collinear factorization~\cite{Lee:2024esz,Lee:2025okn,Guo:2025zwb,Kang:2025zto,Herrmann:2025fqy}. Away from special kinematic regimes, the anomalous scaling of the linear power correction is shown to be universal for a broad class of IRC-safe observables with the D (or decay) scheme~\cite{Farren-Colloty:2025amh}, using a renormalon-inspired gluer approach and soft gluon resummation. While this result provides a strikingly simple evolution structure, its derivation relies on a physically motivated framework whose connection to an operator formalism remains to be established.

In this Letter, we will focus on the power corrections in EEC more theoretically and construct a light-ray operator to describe the leading power correction. Through an explicit one-loop computation in dimensional regularization with $d = 4 - 2\epsilon$, we extract the one-loop anomalous dimension from the $1/\epsilon$ pole and our result is consistent with Refs.~\cite{Dasgupta:2024znl,Farren-Colloty:2025amh,Banfi:2025crj}. Furthermore, we establish a connection with Balitsky–Fadin–Kuraev–Lipatov (BFKL) anomalous dimension~\cite{Kuraev:1977fs, Balitsky:1978ic,Lipatov:1996ts} using the detector matching procedure~\cite{Chang:2025zib,Chang:2025kgq}.

\section{EEC and leading power correction}

EEC, denoted as $\langle \cE(z_1) \cE(z_2)\rangle$, is a correlation function of energy detectors defined as
\be
\cE(z_i) = \lim_{r_i\to \infty} r^2 \int_0^\infty dt\; \vec{n}^j T^{0}_{\; j}(t, r_i \vec{n}_i),,
\ee
where $z_i=(1,\vec{n}_i)$ are null vectors. We consider the simplest case where the source has rotational symmetry in the center-of-mass frame $p=(Q,\vec{0})$, and $p$ denotes the total momentum. In this case, the EEC depends non-trivially only on a single variable $\zeta = \frac{(z_1\cdot z_2) p^2}{2(z_1\cdot p)(z_2\cdot p)}$,
\be
\langle \cE(z_1) \cE(z_2)\rangle = \frac{(p^2)^{d-1}}{(z_1\cdot p)^{d-1} (z_2\cdot p)^{d-1}} \cF(\zeta)\,.
\ee

As an IRC-safe observable, EEC can be well approximated by a perturbative calculation in the large-$Q$ limit, with non-perturbative effects suppressed by powers of $\Lambda_{\QCD}/Q$. The leading power correction contribution scales linearly with $1/Q$, which can be shown using dispersive approach~\cite{Dokshitzer:1999sh}, shape functions~\cite{Korchemsky:1999kt,Belitsky:2001ij} as well as renormalon calculations~\cite{Nason:1995np,Schindler:2023cww}. 
The underlying physical picture is unified across these methods: a non-perturbative soft particle is measured by one of the detectors.
 Formally, this effect is captured by the matrix elements of the energy detector inside Wilson lines, $Y_n = \mathrm{P} \exp\left[{i g\int_0^\infty ds \, n\cdot A(s n)}\right]$:
\begin{align}
\mathcal{G}_{n_1,\dots, n_k} (z;\mu) = \langle  Y^\dagger_{n_1}\cdots Y^\dagger_{n_k} \cE(z) Y_{n_k}\cdots Y_{n_1}\rangle_{\mu}\,, \; 
\end{align}
where $\mu$ denotes the ultraviolet (UV) renormalization scale. In the dijet limit, the dipole configuration ($k=2$) dictates the dominant power correction~\cite{Korchemsky:1999kt,Belitsky:2001ij,Lee:2006nr,Mateu:2012nk}. Even away from the strict dijet limit, this remains the primary source of power corrections at perturbative scales $\mu \gg \Lambda_{\QCD}$, as configurations requiring additional hard Wilson lines are parametrically suppressed by the strong coupling, $\alpha_s(\mu)$.

Suppose the masses of all final-state particles are negligible, $\mathcal{G}_{n_1,n_2} (z;\mu)$ is completely fixed by Lorentz symmetry~\cite{Korchemsky:1999kt,Belitsky:2001ij}
\begin{align}
\mathcal{G}_{n_1,n_2} (z;\mu)
= \Omega(\mu) \left(\frac{(2 n_1 \cdot n_2)}{(2z\cdot n_1)(2z\cdot n_2)}\right)^{\frac{d-1}{2}}\,,
\label{eq:G2-function}
\end{align}
where $\Omega(\mu)\sim \Lambda_\QCD$ is a non-perturbative parameter that is universal for other dijet event shapes in the massless limit. This universality is broken by hadron mass corrections, which depend on the scheme used to reconstruct particle energy and momentum~\cite{Salam:2001bd,Mateu:2012nk}.

\begin{figure}[htbp]
    \centering
    \subfloat[]{
        \includegraphics[width=0.20\textwidth]{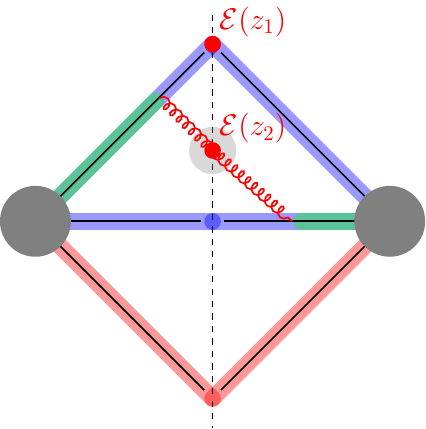}
        \label{fig:sub1}
    }
    \hfill
    \subfloat[]{
        \includegraphics[width=0.2\textwidth]{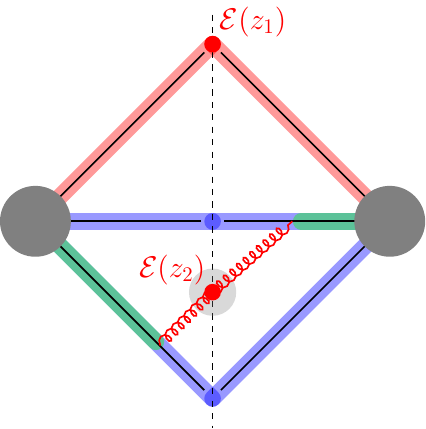}
        \label{fig:sub2}
    }
    \caption{A schematic illustration of the leading power correction of EEC.  $\cE(z_1)$ is placed on a energetic final-state particle, while $\cE(z_2)$ measures a non-perturbative soft particle (represented by the gray blob). The colors on straight legs emphasize the color rotation effect due to soft gluon emission.}
    \label{fig:schematic_diagrams}
\end{figure}

In this Letter, we consider the massless case, {\it e.g.}, by assuming the D scheme, where each hadron decays isotropically into two massless particles.
We propose a light-ray operator framework to describe the EEC that explicitly isolates the leading power correction~\footnote{In this Letter, we neglect the contribution from multi-leg Wilson line contribution that is suppressed by extra powers of $\alpha_s$.}
\begin{align}
\langle \cE(z_1) \cE(z_2)\rangle \simeq \langle \cE(z_1) \cE(z_2)\rangle_{\text{P}} + \frac{\Omega(\mu)}{Q} \langle \cH(z_1,z_2;\mu) \rangle ,
\end{align}
where $\langle \cE(z_1) \cE(z_2)\rangle_{\text{P}}$ represents the perturbative contribution. The light-ray operator $\cH(z_1,z_2;\mu)$ is:
\begin{align}\label{eq:H-components}
\cH(z_1,z_2) = \left[\Hdijet(z_1, z_2) + \Htrijet(z_1, z_2)\right] + (z_1\leftrightarrow z_2)\,,
\end{align}
with dijet and triple-jet operator $\Hdijet, \Htrijet$ given by
\begin{widetext}
\begin{align}
\Hdijet (z_1,z_2) &= \int D^{d-2}z_3 \left(\frac{(2 z_1\cdot z_3)}{(2 z_2\cdot z_1)(2 z_2\cdot z_3)}\right)^{\frac{d-1}{2}}  :\cE^c(z_1) \cN^c(z_3): \,,
\label{eq:D1_def}\\
\Htrijet (z_1, z_2) &=  \frac{1}{2}\int D^{d-2} z_3 D^{d-2} z_4 \left(\frac{(2 z_3\cdot z_4)}{(2 z_2\cdot z_3)(2 z_2\cdot z_4)}\right)^{\frac{d-1}{2}}  :\cE(z_1) \cN^c(z_3) \cN^c(z_4):\,.
\label{eq:D2_def}
\end{align}
\end{widetext}
Here $D^{d-2} z_i$ is the measure on the $(d-2)$-dimensional celestial sphere, $:\,:$ is the normal ordering operation.
$\cE^c=\cD_{1-d}^c$ and $\cN^c=\cD_{2-d}^c$ are special cases of the operator $\cD^c_{J_L}$ defined in~\cite{Chang:2025zib}
\begin{align}\label{eq:color-detector}
\cD^c_{J_L}(z) = \sum_{f}\! \int\! \frac{E^{-J_L}dE}{(2\pi)^{d-1}2E} \mathbf{T}^c_{IJ} a^\dagger_{f, I}(p) a_{f, J}(p)\Big|_{p=E z},
\end{align}
where $f$ labels particle flavors in perturbative QCD, $c$ is an adjoint color index, and $\mathbf{T}^c_{IJ}$ is the color matrix associated with the color representation of $f$, carrying color indices $I,J$. 

The construction of $\cH(z_1, z_2)$ is dual to \eqref{eq:G2-function}, where the kinematic dependence in $\mathcal{G}_{n_1,n_2} (z;\mu)$ becomes the celestial integration kernel in \eqref{eq:D1_def} and \eqref{eq:D2_def}. The underlying physics for $\Hdijet$ and $\Htrijet$ is schematically illustrated in figure \ref{fig:schematic_diagrams}: two cases arise depending on whether the detector $\cE(z_1)$ measures a hard leg of the dipole responsible for emitting the soft particle detected by $\cE(z_2)$. The final-state measurement cannot distinguish these two contributions and hence leads to the combination $\Hdijet+\Htrijet$. This particular combination is important for calculating the anomalous dimension in the next section. 

We further decompose the composite operator $\cH(z_1,z_2)$ into irreducible Lorentz group representations using the diagonalization introduced in~\cite{Caron-Huot:2022eqs}
\begin{align}
\cH_{J_L}(z) = \frac{\Gamma(d-2+J_L)}{\Gamma(\frac{d-2+J_L}{2}){}^2} \int D^{d-2}z_1 D^{d-2}z_2 \qquad
\nn\\
 K_{J_L;1-d,1-d}(z;z_1,z_2) \cH(z_1,z_2)\,,
\end{align}
where $J_L$ is the spin of detector and the kernel $K_{j; j_1, j_2}(z; z_1, z_2)$ is
\begin{align}
K_{j; j_1, j_2}(z; z_1, z_2) = \frac{ (2z_2\cdot z)^{j+j_{1}-j_{2} \over 2} (2z_1\cdot z)^{j+j_{2}-j_{1} \over 2}}{(2z_1\cdot z_2)^{j+j_{1}+j_{2}+2(d-2) \over 2}}.
\end{align}
We define $\HdijetJL(z),\,\HtrijetJL(z)$ in the same way from $\Hdijet(z_1,z_2), \Htrijet(z_1,z_2)$. Due to the $z_1\leftrightarrow z_2$ symmetry in \eqref{eq:H-components}, we have the relation
\begin{align}
\cH_{J_L} (z) = 2\left[\HdijetJL(z)+\HtrijetJL(z)\right]\,.
\end{align}

\section{IR divergences and renormalization}

In this section, we calculate the matrix elements of $\langle \cH_{J_L}(z) \rangle$. Through an explicit one-loop calculation, we extract the one-loop anomalous dimension from IR divergences. For simplicity, we choose $e^+e^-\to \gamma^* \to X$ as the underlying process and integrate out the beam axis dependence.

At tree level, the final state is a two-particle state $|q\bar{q}\rangle$. Inside this state, the matrix element of triple-jet component $\HtrijetJL(z)$ vanishes.
The tree-level result is
\begin{align}
&\langle \cH_{J_L}(z)  \rangle_{\text{tree}} =2 \langle \HdijetJL(z)  \rangle_{\text{tree}}\nn\\
&\; =2C_{J_L,d}\left[ C_F N_c \frac{(2-d) 2^{3-d}\pi^{1-\frac{d}{2}}}{\Gamma(\frac{d-2}{2})}\right] \frac{(2z\cdot p)^{J_L}}{(p^2)^{J_L-d+1 \over 2}}\,,
\end{align}
where the constant $C_{J_L,d}$ is
\begin{align}
 C_{J_L,d} = - \frac{2 \pi^{\frac{d-1}{2}} \Gamma(\frac{1-J_L}{2})}{\Gamma(\frac{d-1}{2})\Gamma(-\frac{J_L}{2})}
\frac{ \Gamma(\frac{d-1+J_L}{2})}{\Gamma(\frac{d-2+J_L}{2})} \,.
\end{align}

At one loop, we separately compute virtual and real corrections. The virtual correction is non-vanishing only for $\langle \HdijetJL(z) \rangle$. In particular, the corresponding $\epsilon$ poles can be read off directly from the universality of IR divergences in one-loop QCD amplitudes~\cite{Giele:1991vf,Kunszt:1994np,Catani:1996vz}
\begin{align}
\langle \HdijetJL(z) \rangle_{\text{1-loop}}^{\text{V}} & = 
\frac{g^2 \tilde{\mu}^{2\epsilon} C_F}{(4\pi)^2} \langle \HdijetJL(z) \rangle_{\text{tree}}
\Big[-\frac{4}{\epsilon^2} 
\nn\\
& +\frac{2}{\epsilon}
\left(2\log \frac{e^{\gamma_E}}{4\pi} p^2 -3 \right)\Big]
+ \mathcal{O}(\epsilon^0)\,.
\label{eq: D1_virtual}
\end{align}

We calculate the real emission contribution to $\langle \HdijetJL(z) \rangle$ by rewriting the correlator:
\begin{align}
& \langle \HdijetJL(z) \rangle = C_{J_L,d} \frac{\Gamma(d-2+J_L)}{\Gamma(\frac{d-3+J_L}{2}) \Gamma(\frac{d-1+J_L}{2})}
\\
&  \int D^{d-2}z_1 D^{d-2}z_3
 K_{J_L; 1-d, 2-d}(z; z_1, z_3)\; 
\langle \cE^c(z_1) \cN^c(z_3) \rangle \,.\nn
\label{eq:op1_calc_def}
\end{align}
This integral falls into the same class as the one evaluated in~\cite{Chang:2025zib}. We follow the same procedure to extract the $\epsilon$ poles from the real emission contribution and combine it with \eqref{eq: D1_virtual}. The full one-loop divergence for $\langle \HdijetJL(z) \rangle$ is
\begin{widetext}
\begin{align}
{\langle \HdijetJL(z) \rangle_{\text{1-loop}} \over \langle \HdijetJL(z) \rangle_{\text{tree}}}
=
\frac{g^2 \tilde{\mu}^{2\epsilon} C_A}{(4\pi)^2} 
\left\{
-\frac{1}{\epsilon^2} 
+ \frac{1}{\epsilon} \left[\psi\left(\frac{3+J_L}{2}\right)+\psi\left(\frac{1-J_L}{2} \right)+3\gamma_E+\log \frac{p^2}{4\pi}
+\frac{8C_F}{3C_A} -\frac{25}{6}
\right]
\right\} +\mathcal{O}(\epsilon^0)\,,
\end{align}
\end{widetext}
where $\psi(x) = \Gamma^\prime(x)/\Gamma(x)$ is the digamma function. We notice that $\langle \HdijetJL(z) \rangle$ has a $1/\epsilon^2$ double pole at one loop, which is not a typical feature of standard operator renormalization. This phenomenon is common in soft-collinear effective theory, which leads to an explicitly scale-dependent anomalous dimension; see, e.g., Refs.~\cite{Becher:2006nr,Becher:2009cu}. In this case, we cannot write down a renormalization-group (RG) equation without referring to a specific source/state.

The second operator, $\langle \HtrijetJL(z) \rangle$, plays an important role in formulating a state-independent RG equation. We first calculate divergences of the matrix element $\langle \Htrijet(z_1, z_2) \rangle$:
\begin{align}
&\langle \Htrijet(z_1, z_2) \rangle_{\text{1-loop}} =  \frac{1}{2} \frac{(p^2)^{\frac{9}{2}-4\epsilon}}{( (z_1\cdot p)(z_2\cdot p))^{3-2\epsilon}} \frac{g^2\tilde{\mu}^{2\epsilon}C_F N_c}{(4\pi)^2}
\nn\\
&\quad \left[
 -\frac{C_A}{16\pi^2\epsilon}
 \frac{\log (1-\zeta)}{(\zeta(1-\zeta))^{3/2}}
  -\frac{C_A }{192\pi^2}  \frac{1}{(\zeta(1-\zeta))^{3/2}}
 \right. \nn\\
&\quad \;\; \left.
 \left( \frac{6}{\epsilon^2} +\frac{43}{\epsilon}
 -\frac{16 }{ \epsilon}\frac{C_F}{C_A}  +\frac{6}{\epsilon}\log \frac{16\pi^3}{e^{\gamma_E}}
 \right)
 \right]+\mathcal{O}(\epsilon^0)\,.
\end{align}
By performing the celestial integration over $z_1, z_2$ against the kernel $ \frac{\Gamma(d-2+J_L)}{\Gamma(\frac{d-2+J_L}{2}){}^2} K_{J_L;1-d,1-d}(z;z_1,z_2)$, we obtain the $\epsilon$ expansion of $\langle \HtrijetJL(z) \rangle$
\begin{widetext}
\begin{align}
{\langle \HtrijetJL(z) \rangle_{\text{1-loop}}
\over \langle \HdijetJL(z) \rangle_{\text{tree}} }
=
\frac{g^2 \tilde{\mu}^{2\epsilon} C_A}{(4\pi)^2}
\left\{
\frac{1}{\epsilon^2} - \frac{1}{\epsilon} \left[
\psi\left(\frac{3+J_L}{2}\right)+\psi\left(\frac{1-J_L}{2} \right)+
3\gamma_E+\log \frac{p^2}{4\pi}+8\log 2
+\frac{8C_F}{3C_A} -\frac{73}{6}
\right]
\right\} +\mathcal{O}(\epsilon)\,.
\end{align}
\end{widetext}
Remarkably, the $1/\epsilon^2$ double poles in $\langle \HdijetJL(z) \rangle_{\text{1-loop}} $ and $\langle \HtrijetJL(z) \rangle_{\text{1-loop}}$ exactly cancel, leaving a standard single-pole structure
\begin{align}
\langle \cH_{J_L}(z) \rangle_{\text{1-loop}} = \frac{g^2 C_A}{(4\pi)^2\epsilon} \cS_1 \langle \cH_{J_L}(z) \rangle_{\text{tree}} +\mathcal{O}(\epsilon^0)\,,
\end{align}
where the constant $\cS_1 = 8 (1-\log 2)$.
This indicates that the particular combination of $\HdijetJL(z)$ and $\HtrijetJL(z)$ makes it important to define a state-independent RG equation.
 
 Following the standard renormalization procedure in the $\overline{\mathrm{MS}}$ scheme, we define the renormalized operator $[\cH_{J_L}]_R(z)$ by introducing the renormalization factor $\mathcal{Z}_{J_L}$
\begin{align}
\cH_{J_L}(z) &= \mathcal{Z}_{J_L} [\cH_{J_L}]_R(z)\,,\\
 \mathcal{Z}_{J_L}  &= 1+\frac{\alpha_s C_A}{4\pi\epsilon}\cS_1 +\mathcal{O}(\alpha_s^2)\,.
\end{align}
This leads to the RG equation for $\cH_{J_L}$
\begin{align}\label{eq: RGE}
 \mu \frac{d}{d\mu} [\cH_{J_L}]_R(z;\mu) = \frac{\alpha_s C_A}{2\pi} \cS_1   [\cH_{J_L}]_R(z;\mu)
 +\mathcal{O}(\alpha_s^2)\,,
\end{align}
whose matrix element has the following leading-logarithmic solution
\begin{align}
\langle [\cH_{J_L}]_R(z;\mu)\rangle = \langle [\cH_{J_L}]_R(z;Q)\rangle \left(\frac{\alpha_s(Q)}{\alpha_s(\mu)}\right)^{\frac{C_A \cS_1}{\beta_0}}\,.
\end{align}
Since the $\cS$ constant is independent of $J_L$, we can immediately conclude that $\langle \cH(z_1,z_2)\rangle$ satisfies the same relations.
This reproduces the same leading-logarithmic structure found in~\cite{Farren-Colloty:2025amh} for the leading power correction to the EEC.

\begin{figure}[htbp]
    \centering
    \includegraphics[width=0.4\textwidth]{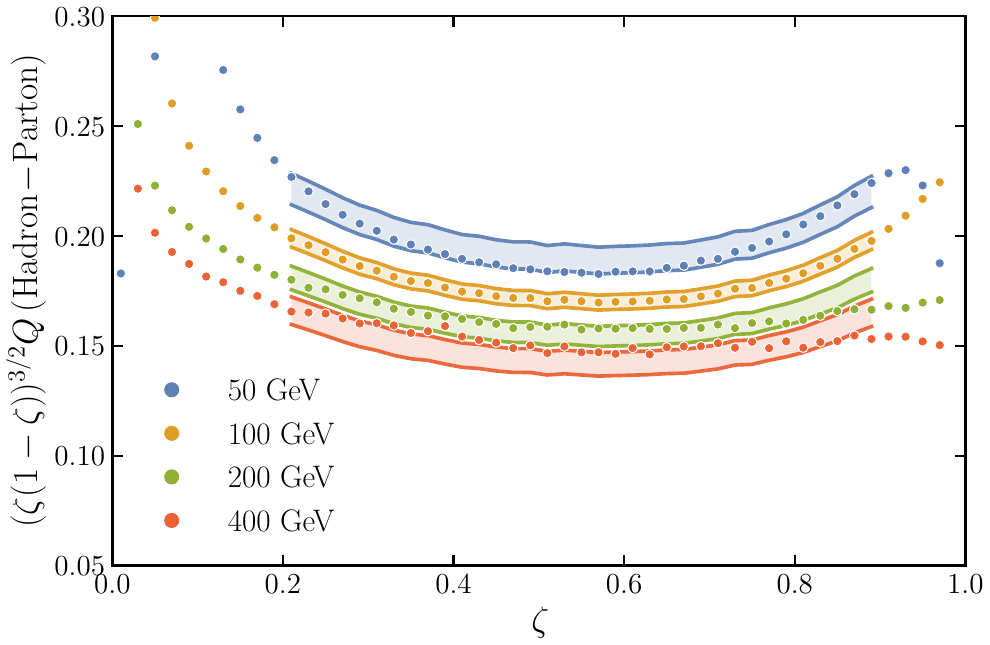}
    \caption{
    Comparison of the leading power correction to the EEC extracted from Pythia simulations (dots) and our theoretical RG evolution (bands). The analytical prediction uses the Pythia result at $100\,\mathrm{GeV}$ as the initial condition, with the bands representing the scale uncertainty.
    }
    \label{fig:pythia-plot}
\end{figure}

To validate this evolution equation, we implement the D scheme within the Pythia event generator~\cite{Bierlich:2022pfr}. In Figure~\ref{fig:pythia-plot}, we show the isolated leading power correction, defined as the difference between the hadron- and parton-level EECs scaled by Q:
\begin{align}
(\zeta(1-\zeta))^{\frac{3}{2}}Q\left[\cF_Q^{\text{D-scheme}}(\zeta) - \cF_Q^{\text{parton}}(\zeta)\right]\,,
\end{align}
This is evaluated at several center-of-mass energies $Q=50,100,200,400\, \mathrm{GeV}$.
Taking $Q_0=100\, \mathrm{GeV}$ as our initial condition and assigning a conservative $2\%$ uncertainty band, we apply the analytic evolution kernel
$  \left(\alpha_s(\xi Q)/\alpha_s(\xi Q_0)\right)^{\frac{C_A \cS_1}{\beta_0}}$
to predict the distributions at the remaining energies. The standard scale variation $0.5<\xi<2$ is used to estimate the theoretical uncertainty of the evolution.
We observe good agreement between our analytical evolution and the Pythia simulations within the bulk kinematic region, $0.2<\zeta<0.9$. 
Near the endpoints ($\zeta\sim 0$ and $\zeta\sim 1$), we expect other dynamics—including collinear physics—to become important~\cite{Chen:2024nyc, Lee:2024esz}, which is beyond the scope of this Letter.

\section{Connection to BFKL physics}

We notice an interesting relation between the leading-order (LO) anomalous dimension in \eqref{eq: RGE} and LO BFKL anomalous dimension. 
BFKL anomalous dimension is the eigenvalue of the BFKL equation~\cite{Kuraev:1977fs,Balitsky:1978ic}. We follow the notation in~\cite{Chang:2025zib}, where the LO anomalous dimension is 
\begin{align}\label{eq:gamma_BFKL}
\gamma_{\BFKL}(J_L) = \frac{\alpha_s}{2\pi} C_A \left(2\gamma_E+\psi(\tfrac{2+J_L}{2}) + \psi(-\tfrac{J_L}{2})\right)\,,
\end{align}
By setting $J_L= -3$, we find the relation
\begin{align}\label{eq:gamma--3}
\gamma_{\BFKL}(J_L=-3) = \frac{\alpha_s C_A}{4\pi} \cS_1\,.
\end{align}

This relation can be understood by the detector matching~\cite{Chang:2025zib,Chang:2025kgq}. In a large-$Q$ state,  the matching procedure expands a non-perturbative detector onto a sum of perturbative detectors. The matching equation for $\cE$ is
\begin{align}\label{eq: E-matching}
\mathcal{E}(z) \simeq \mathcal{E}_{\text{pert.}}(z) + \Omega\, \cD^\BFKL_{J_L=1-d}(z) + \cdots\,,
\end{align}
where $\cE_{\text{pert.}}$ represents the perturbative energy detector written in terms of quark and gluon, $\cD^\BFKL_{J_L}(z)$ is the BFKL detector defined in~\cite{Chang:2025zib}
\begin{align}
&\cD_{J_L}^\BFKL(z) \propto \\
&\frac{1}{2}\int D^{d-2}z_3 D^{d-2} z_4  K_{J_L; 1-d,1-d}(z;z_3,z_4):\cN^c(z_3) \cN^c(z_4):
\nn
\end{align}
whose LO anomalous dimension is \eqref{eq:gamma_BFKL}. This matching procedure is the generalization of the collinear fragmentation picture: for instance, the power correction term in \eqref{eq: E-matching} is related to soft emissions.

Interestingly, we notice the combination $\Hdijet(z_1, z_2) + \Htrijet(z_1, z_2)$ in \eqref{eq:H-components} follows directly from the normal ordering procedure in free theory
\begin{align}
\cE(z_1) \cD^\BFKL_{J_L=1-d}(z_2) \propto \Hdijet(z_1, z_2) + \Htrijet(z_1, z_2)\,,
\end{align}
because of the relation following from \eqref{eq:color-detector}:
\begin{align}
&\cE(z_1) :\cN^c(z_3) \cN^c(z_4): \;\;
= 
\;\; :\cE(z_1)\cN^c(z_3) \cN^c(z_4): 
 \nn\\
&\qquad \quad + 
\left[
\delta(z_1, z_3) :\cE^c(z_1) \cN^c(z_4):
+(z_3\leftrightarrow z_4)\right] \,,
\end{align}
where $\delta(z_i, z_j) $ is $(d-2)$-dimensional delta function on the celestial sphere.

Essentially, it is the anomalous dimension of $\cD^\BFKL_{J_L=1-d}(z) $, given by \eqref{eq:gamma--3}, that controls the scale evolution of leading EEC power correction.
This is straightforward if $\cD^\BFKL_{J_L=1-d}(z)$ is localized on the celestial sphere.
However, the subtlety is that the definition of $\cD^\BFKL_{J_L}(z)$ is non-local on the celestial sphere. It is not evident whether extra insertion of measurements can change the IR divergences. The explicit computation in the previous section confirms that an extra insertion of $\cE$ does not introduce additional IR divergences. In other words, $\cE(z) \cD^\BFKL_{J_L=1-d} (z^\prime)$ is a trivial composite detector in the sense that two operators renormalize independently. 
We conjecture that this property is general for light-ray operators above Regge intercept~\cite{Kologlu:2019bco,Kologlu:2019mfz}.

\section{Conclusion and Outlook}

In this Letter, we developed a light-ray operator framework to investigate the leading power correction to the EEC and its associated $Q$-evolution at generic angles, away from the endpoint regions. By explicitly constructing the relevant light-ray operator, we recast the anomalous scaling of the power correction as a rigorous operator renormalization procedure in quantum field theory. We computed the one-loop anomalous dimension and, via a detector matching procedure, identified its physical origin in BFKL dynamics. Finally, we demonstrated that the analytically predicted anomalous scaling is in good agreement with Pythia simulations employing the D scheme.

Our work opens several exciting avenues for future theoretical research. The most immediate goal is to establish a smooth connection to the endpoint regions, where an all-orders resummation of the relevant kinematic parameters is required. Additional physical modes become relevant, such as the collinear dynamics discussed in~\cite{Chen:2024nyc,Lee:2024esz}. Second, extending our one-loop calculations to higher orders and applying this operator framework to hadronic initial states will be critical steps toward achieving unprecedented theoretical precision for modern collider phenomenology. Furthermore, it will be highly instructive to investigate other hadron mass schemes within the light-ray operator framework. From a purely formal perspective, systematically mapping the full operator space of light-ray operators remains a profound challenge—one that promises to provide a universal, rigorous language for interesting QCD phenomenology at the high-energy frontier.

\begin{acknowledgments}
We thank Cyuan-Han Chang, Pier Monni, David Simmons-Duffin, Iain Stewart, Xiaoyuan Zhang and Hua Xing Zhu for helpful discussions.
H.C. is supported by the U.S. Department of Energy, Office of Science, Office of Nuclear Physics under grant Contract Number DESC0011090. 
Y.L. is supported by funding from the European Research Council (ERC) under the European Union's Horizon 2022 Research and Innovation Program (ERC Advanced Grant agreement No.101097780, EFT4jets).
\end{acknowledgments}

\bibliography{refs}%

\begin{thebibliography}{66}%
\makeatletter
\providecommand \@ifxundefined [1]{%
 \@ifx{#1\undefined}
}%
\providecommand \@ifnum [1]{%
 \ifnum #1\expandafter \@firstoftwo
 \else \expandafter \@secondoftwo
 \fi
}%
\providecommand \@ifx [1]{%
 \ifx #1\expandafter \@firstoftwo
 \else \expandafter \@secondoftwo
 \fi
}%
\providecommand \natexlab [1]{#1}%
\providecommand \enquote  [1]{``#1''}%
\providecommand \bibnamefont  [1]{#1}%
\providecommand \bibfnamefont [1]{#1}%
\providecommand \citenamefont [1]{#1}%
\providecommand \href@noop [0]{\@secondoftwo}%
\providecommand \href [0]{\begingroup \@sanitize@url \@href}%
\providecommand \@href[1]{\@@startlink{#1}\@@href}%
\providecommand \@@href[1]{\endgroup#1\@@endlink}%
\providecommand \@sanitize@url [0]{\catcode `\\12\catcode `\$12\catcode
  `\&12\catcode `\#12\catcode `\^12\catcode `\_12\catcode `\%12\relax}%
\providecommand \@@startlink[1]{}%
\providecommand \@@endlink[0]{}%
\providecommand \url  [0]{\begingroup\@sanitize@url \@url }%
\providecommand \@url [1]{\endgroup\@href {#1}{\urlprefix }}%
\providecommand \urlprefix  [0]{URL }%
\providecommand \Eprint [0]{\href }%
\providecommand \doibase [0]{https://doi.org/}%
\providecommand \selectlanguage [0]{\@gobble}%
\providecommand \bibinfo  [0]{\@secondoftwo}%
\providecommand \bibfield  [0]{\@secondoftwo}%
\providecommand \translation [1]{[#1]}%
\providecommand \BibitemOpen [0]{}%
\providecommand \bibitemStop [0]{}%
\providecommand \bibitemNoStop [0]{.\EOS\space}%
\providecommand \EOS [0]{\spacefactor3000\relax}%
\providecommand \BibitemShut  [1]{\csname bibitem#1\endcsname}%
\let\auto@bib@innerbib\@empty
\bibitem [{\citenamefont {Basham}\ \emph
  {et~al.}(1978{\natexlab{a}})\citenamefont {Basham}, \citenamefont {Brown},
  \citenamefont {Ellis},\ and\ \citenamefont {Love}}]{Basham:1978bw}%
  \BibitemOpen
  \bibfield  {author} {\bibinfo {author} {\bibfnamefont {C.~L.}\ \bibnamefont
  {Basham}}, \bibinfo {author} {\bibfnamefont {L.~S.}\ \bibnamefont {Brown}},
  \bibinfo {author} {\bibfnamefont {S.~D.}\ \bibnamefont {Ellis}},\ and\
  \bibinfo {author} {\bibfnamefont {S.~T.}\ \bibnamefont {Love}},\ }\bibfield
  {title} {\bibinfo {title} {{Energy Correlations in electron - Positron
  Annihilation: Testing QCD}},\ }\href
  {https://doi.org/10.1103/PhysRevLett.41.1585} {\bibfield  {journal} {\bibinfo
   {journal} {Phys. Rev. Lett.}\ }\textbf {\bibinfo {volume} {41}},\ \bibinfo
  {pages} {1585} (\bibinfo {year} {1978}{\natexlab{a}})}\BibitemShut {NoStop}%
\bibitem [{\citenamefont {Basham}\ \emph
  {et~al.}(1978{\natexlab{b}})\citenamefont {Basham}, \citenamefont {Brown},
  \citenamefont {Ellis},\ and\ \citenamefont {Love}}]{Basham:1977iq}%
  \BibitemOpen
  \bibfield  {author} {\bibinfo {author} {\bibfnamefont {C.~L.}\ \bibnamefont
  {Basham}}, \bibinfo {author} {\bibfnamefont {L.~S.}\ \bibnamefont {Brown}},
  \bibinfo {author} {\bibfnamefont {S.~D.}\ \bibnamefont {Ellis}},\ and\
  \bibinfo {author} {\bibfnamefont {S.~T.}\ \bibnamefont {Love}},\ }\bibfield
  {title} {\bibinfo {title} {{Electron - Positron Annihilation Energy Pattern
  in Quantum Chromodynamics: Asymptotically Free Perturbation Theory}},\ }\href
  {https://doi.org/10.1103/PhysRevD.17.2298} {\bibfield  {journal} {\bibinfo
  {journal} {Phys. Rev. D}\ }\textbf {\bibinfo {volume} {17}},\ \bibinfo
  {pages} {2298} (\bibinfo {year} {1978}{\natexlab{b}})}\BibitemShut {NoStop}%
\bibitem [{\citenamefont {Basham}\ \emph
  {et~al.}(1979{\natexlab{a}})\citenamefont {Basham}, \citenamefont {Brown},
  \citenamefont {Ellis},\ and\ \citenamefont {Love}}]{Basham:1979gh}%
  \BibitemOpen
  \bibfield  {author} {\bibinfo {author} {\bibfnamefont {C.~L.}\ \bibnamefont
  {Basham}}, \bibinfo {author} {\bibfnamefont {L.~S.}\ \bibnamefont {Brown}},
  \bibinfo {author} {\bibfnamefont {S.~D.}\ \bibnamefont {Ellis}},\ and\
  \bibinfo {author} {\bibfnamefont {S.~T.}\ \bibnamefont {Love}},\ }\bibfield
  {title} {\bibinfo {title} {{Energy Correlations in Perturbative Quantum
  Chromodynamics: A Conjecture for All Orders}},\ }\href
  {https://doi.org/10.1016/0370-2693(79)90601-4} {\bibfield  {journal}
  {\bibinfo  {journal} {Phys. Lett. B}\ }\textbf {\bibinfo {volume} {85}},\
  \bibinfo {pages} {297} (\bibinfo {year} {1979}{\natexlab{a}})}\BibitemShut
  {NoStop}%
\bibitem [{\citenamefont {Basham}\ \emph
  {et~al.}(1979{\natexlab{b}})\citenamefont {Basham}, \citenamefont {Brown},
  \citenamefont {Ellis},\ and\ \citenamefont {Love}}]{Basham:1978zq}%
  \BibitemOpen
  \bibfield  {author} {\bibinfo {author} {\bibfnamefont {C.~L.}\ \bibnamefont
  {Basham}}, \bibinfo {author} {\bibfnamefont {L.~S.}\ \bibnamefont {Brown}},
  \bibinfo {author} {\bibfnamefont {S.~D.}\ \bibnamefont {Ellis}},\ and\
  \bibinfo {author} {\bibfnamefont {S.~T.}\ \bibnamefont {Love}},\ }\bibfield
  {title} {\bibinfo {title} {{Energy Correlations in electron-Positron
  Annihilation in Quantum Chromodynamics: Asymptotically Free Perturbation
  Theory}},\ }\href {https://doi.org/10.1103/PhysRevD.19.2018} {\bibfield
  {journal} {\bibinfo  {journal} {Phys. Rev. D}\ }\textbf {\bibinfo {volume}
  {19}},\ \bibinfo {pages} {2018} (\bibinfo {year}
  {1979}{\natexlab{b}})}\BibitemShut {NoStop}%
\bibitem [{\citenamefont {Moult}\ and\ \citenamefont
  {Zhu}(2025)}]{Moult:2025nhu}%
  \BibitemOpen
  \bibfield  {author} {\bibinfo {author} {\bibfnamefont {I.}~\bibnamefont
  {Moult}}\ and\ \bibinfo {author} {\bibfnamefont {H.~X.}\ \bibnamefont
  {Zhu}},\ }\bibfield  {title} {\bibinfo {title} {{Energy Correlators: A
  Journey From Theory to Experiment}},\ }\href@noop {} {\  (\bibinfo {year}
  {2025})},\ \Eprint {https://arxiv.org/abs/2506.09119} {arXiv:2506.09119
  [hep-ph]} \BibitemShut {NoStop}%
\bibitem [{\citenamefont {Hofman}\ and\ \citenamefont
  {Maldacena}(2008)}]{Hofman:2008ar}%
  \BibitemOpen
  \bibfield  {author} {\bibinfo {author} {\bibfnamefont {D.~M.}\ \bibnamefont
  {Hofman}}\ and\ \bibinfo {author} {\bibfnamefont {J.}~\bibnamefont
  {Maldacena}},\ }\bibfield  {title} {\bibinfo {title} {{Conformal collider
  physics: Energy and charge correlations}},\ }\href
  {https://doi.org/10.1088/1126-6708/2008/05/012} {\bibfield  {journal}
  {\bibinfo  {journal} {JHEP}\ }\textbf {\bibinfo {volume} {05}},\ \bibinfo
  {pages} {012}},\ \Eprint {https://arxiv.org/abs/0803.1467} {arXiv:0803.1467
  [hep-th]} \BibitemShut {NoStop}%
\bibitem [{\citenamefont {Kravchuk}\ and\ \citenamefont
  {Simmons-Duffin}(2018)}]{Kravchuk:2018htv}%
  \BibitemOpen
  \bibfield  {author} {\bibinfo {author} {\bibfnamefont {P.}~\bibnamefont
  {Kravchuk}}\ and\ \bibinfo {author} {\bibfnamefont {D.}~\bibnamefont
  {Simmons-Duffin}},\ }\bibfield  {title} {\bibinfo {title} {{Light-ray
  operators in conformal field theory}},\ }\href
  {https://doi.org/10.1007/JHEP11(2018)102} {\bibfield  {journal} {\bibinfo
  {journal} {JHEP}\ }\textbf {\bibinfo {volume} {11}},\ \bibinfo {pages}
  {102}},\ \Eprint {https://arxiv.org/abs/1805.00098} {arXiv:1805.00098
  [hep-th]} \BibitemShut {NoStop}%
\bibitem [{\citenamefont {Kologlu}\ \emph {et~al.}(2021)\citenamefont
  {Kologlu}, \citenamefont {Kravchuk}, \citenamefont {Simmons-Duffin},\ and\
  \citenamefont {Zhiboedov}}]{Kologlu:2019mfz}%
  \BibitemOpen
  \bibfield  {author} {\bibinfo {author} {\bibfnamefont {M.}~\bibnamefont
  {Kologlu}}, \bibinfo {author} {\bibfnamefont {P.}~\bibnamefont {Kravchuk}},
  \bibinfo {author} {\bibfnamefont {D.}~\bibnamefont {Simmons-Duffin}},\ and\
  \bibinfo {author} {\bibfnamefont {A.}~\bibnamefont {Zhiboedov}},\ }\bibfield
  {title} {\bibinfo {title} {{The light-ray OPE and conformal colliders}},\
  }\href {https://doi.org/10.1007/JHEP01(2021)128} {\bibfield  {journal}
  {\bibinfo  {journal} {JHEP}\ }\textbf {\bibinfo {volume} {01}},\ \bibinfo
  {pages} {128}},\ \Eprint {https://arxiv.org/abs/1905.01311} {arXiv:1905.01311
  [hep-th]} \BibitemShut {NoStop}%
\bibitem [{\citenamefont {Chang}\ \emph {et~al.}(2022)\citenamefont {Chang},
  \citenamefont {Kologlu}, \citenamefont {Kravchuk}, \citenamefont
  {Simmons-Duffin},\ and\ \citenamefont {Zhiboedov}}]{Chang:2020qpj}%
  \BibitemOpen
  \bibfield  {author} {\bibinfo {author} {\bibfnamefont {C.-H.}\ \bibnamefont
  {Chang}}, \bibinfo {author} {\bibfnamefont {M.}~\bibnamefont {Kologlu}},
  \bibinfo {author} {\bibfnamefont {P.}~\bibnamefont {Kravchuk}}, \bibinfo
  {author} {\bibfnamefont {D.}~\bibnamefont {Simmons-Duffin}},\ and\ \bibinfo
  {author} {\bibfnamefont {A.}~\bibnamefont {Zhiboedov}},\ }\bibfield  {title}
  {\bibinfo {title} {{Transverse spin in the light-ray OPE}},\ }\href
  {https://doi.org/10.1007/JHEP05(2022)059} {\bibfield  {journal} {\bibinfo
  {journal} {JHEP}\ }\textbf {\bibinfo {volume} {05}},\ \bibinfo {pages}
  {059}},\ \Eprint {https://arxiv.org/abs/2010.04726} {arXiv:2010.04726
  [hep-th]} \BibitemShut {NoStop}%
\bibitem [{\citenamefont {Chen}\ \emph {et~al.}(2021)\citenamefont {Chen},
  \citenamefont {Moult},\ and\ \citenamefont {Zhu}}]{Chen:2020adz}%
  \BibitemOpen
  \bibfield  {author} {\bibinfo {author} {\bibfnamefont {H.}~\bibnamefont
  {Chen}}, \bibinfo {author} {\bibfnamefont {I.}~\bibnamefont {Moult}},\ and\
  \bibinfo {author} {\bibfnamefont {H.~X.}\ \bibnamefont {Zhu}},\ }\bibfield
  {title} {\bibinfo {title} {{Quantum Interference in Jet Substructure from
  Spinning Gluons}},\ }\href {https://doi.org/10.1103/PhysRevLett.126.112003}
  {\bibfield  {journal} {\bibinfo  {journal} {Phys. Rev. Lett.}\ }\textbf
  {\bibinfo {volume} {126}},\ \bibinfo {pages} {112003} (\bibinfo {year}
  {2021})},\ \Eprint {https://arxiv.org/abs/2011.02492} {arXiv:2011.02492
  [hep-ph]} \BibitemShut {NoStop}%
\bibitem [{\citenamefont {Chen}\ \emph {et~al.}(2022)\citenamefont {Chen},
  \citenamefont {Moult}, \citenamefont {Sandor},\ and\ \citenamefont
  {Zhu}}]{Chen:2022jhb}%
  \BibitemOpen
  \bibfield  {author} {\bibinfo {author} {\bibfnamefont {H.}~\bibnamefont
  {Chen}}, \bibinfo {author} {\bibfnamefont {I.}~\bibnamefont {Moult}},
  \bibinfo {author} {\bibfnamefont {J.}~\bibnamefont {Sandor}},\ and\ \bibinfo
  {author} {\bibfnamefont {H.~X.}\ \bibnamefont {Zhu}},\ }\bibfield  {title}
  {\bibinfo {title} {{Celestial blocks and transverse spin in the three-point
  energy correlator}},\ }\href {https://doi.org/10.1007/JHEP09(2022)199}
  {\bibfield  {journal} {\bibinfo  {journal} {JHEP}\ }\textbf {\bibinfo
  {volume} {09}},\ \bibinfo {pages} {199}},\ \Eprint
  {https://arxiv.org/abs/2202.04085} {arXiv:2202.04085 [hep-ph]} \BibitemShut
  {NoStop}%
\bibitem [{\citenamefont {Chang}\ and\ \citenamefont
  {Simmons-Duffin}(2023)}]{Chang:2022ryc}%
  \BibitemOpen
  \bibfield  {author} {\bibinfo {author} {\bibfnamefont {C.-H.}\ \bibnamefont
  {Chang}}\ and\ \bibinfo {author} {\bibfnamefont {D.}~\bibnamefont
  {Simmons-Duffin}},\ }\bibfield  {title} {\bibinfo {title} {{Three-point
  energy correlators and the celestial block expansion}},\ }\href
  {https://doi.org/10.1007/JHEP02(2023)126} {\bibfield  {journal} {\bibinfo
  {journal} {JHEP}\ }\textbf {\bibinfo {volume} {02}},\ \bibinfo {pages}
  {126}},\ \Eprint {https://arxiv.org/abs/2202.04090} {arXiv:2202.04090
  [hep-th]} \BibitemShut {NoStop}%
\bibitem [{\citenamefont {Komiske}\ \emph {et~al.}(2023)\citenamefont
  {Komiske}, \citenamefont {Moult}, \citenamefont {Thaler},\ and\ \citenamefont
  {Zhu}}]{Komiske:2022enw}%
  \BibitemOpen
  \bibfield  {author} {\bibinfo {author} {\bibfnamefont {P.~T.}\ \bibnamefont
  {Komiske}}, \bibinfo {author} {\bibfnamefont {I.}~\bibnamefont {Moult}},
  \bibinfo {author} {\bibfnamefont {J.}~\bibnamefont {Thaler}},\ and\ \bibinfo
  {author} {\bibfnamefont {H.~X.}\ \bibnamefont {Zhu}},\ }\bibfield  {title}
  {\bibinfo {title} {{Analyzing N-Point Energy Correlators inside Jets with CMS
  Open Data}},\ }\href {https://doi.org/10.1103/PhysRevLett.130.051901}
  {\bibfield  {journal} {\bibinfo  {journal} {Phys. Rev. Lett.}\ }\textbf
  {\bibinfo {volume} {130}},\ \bibinfo {pages} {051901} (\bibinfo {year}
  {2023})},\ \Eprint {https://arxiv.org/abs/2201.07800} {arXiv:2201.07800
  [hep-ph]} \BibitemShut {NoStop}%
\bibitem [{\citenamefont {Del~Duca}\ \emph {et~al.}(2016)\citenamefont
  {Del~Duca}, \citenamefont {Duhr}, \citenamefont {Kardos}, \citenamefont
  {Somogyi},\ and\ \citenamefont {Tr{\'o}cs{\'a}nyi}}]{DelDuca:2016csb}%
  \BibitemOpen
  \bibfield  {author} {\bibinfo {author} {\bibfnamefont {V.}~\bibnamefont
  {Del~Duca}}, \bibinfo {author} {\bibfnamefont {C.}~\bibnamefont {Duhr}},
  \bibinfo {author} {\bibfnamefont {A.}~\bibnamefont {Kardos}}, \bibinfo
  {author} {\bibfnamefont {G.}~\bibnamefont {Somogyi}},\ and\ \bibinfo {author}
  {\bibfnamefont {Z.}~\bibnamefont {Tr{\'o}cs{\'a}nyi}},\ }\bibfield  {title}
  {\bibinfo {title} {{Three-Jet Production in Electron-Positron Collisions at
  Next-to-Next-to-Leading Order Accuracy}},\ }\href
  {https://doi.org/10.1103/PhysRevLett.117.152004} {\bibfield  {journal}
  {\bibinfo  {journal} {Phys. Rev. Lett.}\ }\textbf {\bibinfo {volume} {117}},\
  \bibinfo {pages} {152004} (\bibinfo {year} {2016})},\ \Eprint
  {https://arxiv.org/abs/1603.08927} {arXiv:1603.08927 [hep-ph]} \BibitemShut
  {NoStop}%
\bibitem [{\citenamefont {Tulip{\'a}nt}\ \emph {et~al.}(2017)\citenamefont
  {Tulip{\'a}nt}, \citenamefont {Kardos},\ and\ \citenamefont
  {Somogyi}}]{Tulipant:2017ybb}%
  \BibitemOpen
  \bibfield  {author} {\bibinfo {author} {\bibfnamefont {Z.}~\bibnamefont
  {Tulip{\'a}nt}}, \bibinfo {author} {\bibfnamefont {A.}~\bibnamefont
  {Kardos}},\ and\ \bibinfo {author} {\bibfnamefont {G.}~\bibnamefont
  {Somogyi}},\ }\bibfield  {title} {\bibinfo {title}
  {{Energy{\textendash}energy correlation in electron{\textendash}positron
  annihilation at NNLL + NNLO accuracy}},\ }\href
  {https://doi.org/10.1140/epjc/s10052-017-5320-9} {\bibfield  {journal}
  {\bibinfo  {journal} {Eur. Phys. J. C}\ }\textbf {\bibinfo {volume} {77}},\
  \bibinfo {pages} {749} (\bibinfo {year} {2017})},\ \Eprint
  {https://arxiv.org/abs/1708.04093} {arXiv:1708.04093 [hep-ph]} \BibitemShut
  {NoStop}%
\bibitem [{\citenamefont {Dixon}\ \emph {et~al.}(2018)\citenamefont {Dixon},
  \citenamefont {Luo}, \citenamefont {Shtabovenko}, \citenamefont {Yang},\ and\
  \citenamefont {Zhu}}]{Dixon:2018qgp}%
  \BibitemOpen
  \bibfield  {author} {\bibinfo {author} {\bibfnamefont {L.~J.}\ \bibnamefont
  {Dixon}}, \bibinfo {author} {\bibfnamefont {M.-X.}\ \bibnamefont {Luo}},
  \bibinfo {author} {\bibfnamefont {V.}~\bibnamefont {Shtabovenko}}, \bibinfo
  {author} {\bibfnamefont {T.-Z.}\ \bibnamefont {Yang}},\ and\ \bibinfo
  {author} {\bibfnamefont {H.~X.}\ \bibnamefont {Zhu}},\ }\bibfield  {title}
  {\bibinfo {title} {{Analytical Computation of Energy-Energy Correlation at
  Next-to-Leading Order in QCD}},\ }\href
  {https://doi.org/10.1103/PhysRevLett.120.102001} {\bibfield  {journal}
  {\bibinfo  {journal} {Phys. Rev. Lett.}\ }\textbf {\bibinfo {volume} {120}},\
  \bibinfo {pages} {102001} (\bibinfo {year} {2018})},\ \Eprint
  {https://arxiv.org/abs/1801.03219} {arXiv:1801.03219 [hep-ph]} \BibitemShut
  {NoStop}%
\bibitem [{\citenamefont {Dixon}\ \emph {et~al.}(2019)\citenamefont {Dixon},
  \citenamefont {Moult},\ and\ \citenamefont {Zhu}}]{Dixon:2019uzg}%
  \BibitemOpen
  \bibfield  {author} {\bibinfo {author} {\bibfnamefont {L.~J.}\ \bibnamefont
  {Dixon}}, \bibinfo {author} {\bibfnamefont {I.}~\bibnamefont {Moult}},\ and\
  \bibinfo {author} {\bibfnamefont {H.~X.}\ \bibnamefont {Zhu}},\ }\bibfield
  {title} {\bibinfo {title} {{Collinear limit of the energy-energy
  correlator}},\ }\href {https://doi.org/10.1103/PhysRevD.100.014009}
  {\bibfield  {journal} {\bibinfo  {journal} {Phys. Rev. D}\ }\textbf {\bibinfo
  {volume} {100}},\ \bibinfo {pages} {014009} (\bibinfo {year} {2019})},\
  \Eprint {https://arxiv.org/abs/1905.01310} {arXiv:1905.01310 [hep-ph]}
  \BibitemShut {NoStop}%
\bibitem [{\citenamefont {Moult}\ and\ \citenamefont
  {Zhu}(2018)}]{Moult:2018jzp}%
  \BibitemOpen
  \bibfield  {author} {\bibinfo {author} {\bibfnamefont {I.}~\bibnamefont
  {Moult}}\ and\ \bibinfo {author} {\bibfnamefont {H.~X.}\ \bibnamefont
  {Zhu}},\ }\bibfield  {title} {\bibinfo {title} {{Simplicity from Recoil: The
  Three-Loop Soft Function and Factorization for the Energy-Energy
  Correlation}},\ }\href {https://doi.org/10.1007/JHEP08(2018)160} {\bibfield
  {journal} {\bibinfo  {journal} {JHEP}\ }\textbf {\bibinfo {volume} {08}},\
  \bibinfo {pages} {160}},\ \Eprint {https://arxiv.org/abs/1801.02627}
  {arXiv:1801.02627 [hep-ph]} \BibitemShut {NoStop}%
\bibitem [{\citenamefont {Duhr}\ \emph {et~al.}(2022)\citenamefont {Duhr},
  \citenamefont {Mistlberger},\ and\ \citenamefont {Vita}}]{Duhr:2022yyp}%
  \BibitemOpen
  \bibfield  {author} {\bibinfo {author} {\bibfnamefont {C.}~\bibnamefont
  {Duhr}}, \bibinfo {author} {\bibfnamefont {B.}~\bibnamefont {Mistlberger}},\
  and\ \bibinfo {author} {\bibfnamefont {G.}~\bibnamefont {Vita}},\ }\bibfield
  {title} {\bibinfo {title} {{Four-Loop Rapidity Anomalous Dimension and Event
  Shapes to Fourth Logarithmic Order}},\ }\href
  {https://doi.org/10.1103/PhysRevLett.129.162001} {\bibfield  {journal}
  {\bibinfo  {journal} {Phys. Rev. Lett.}\ }\textbf {\bibinfo {volume} {129}},\
  \bibinfo {pages} {162001} (\bibinfo {year} {2022})},\ \Eprint
  {https://arxiv.org/abs/2205.02242} {arXiv:2205.02242 [hep-ph]} \BibitemShut
  {NoStop}%
\bibitem [{\citenamefont {Dokshitzer}\ and\ \citenamefont
  {Webber}(1995)}]{Dokshitzer:1995zt}%
  \BibitemOpen
  \bibfield  {author} {\bibinfo {author} {\bibfnamefont {Y.~L.}\ \bibnamefont
  {Dokshitzer}}\ and\ \bibinfo {author} {\bibfnamefont {B.~R.}\ \bibnamefont
  {Webber}},\ }\bibfield  {title} {\bibinfo {title} {{Calculation of power
  corrections to hadronic event shapes}},\ }\href
  {https://doi.org/10.1016/0370-2693(95)00548-Y} {\bibfield  {journal}
  {\bibinfo  {journal} {Phys. Lett. B}\ }\textbf {\bibinfo {volume} {352}},\
  \bibinfo {pages} {451} (\bibinfo {year} {1995})},\ \Eprint
  {https://arxiv.org/abs/hep-ph/9504219} {arXiv:hep-ph/9504219} \BibitemShut
  {NoStop}%
\bibitem [{\citenamefont {Akhoury}\ and\ \citenamefont
  {Zakharov}(1995)}]{Akhoury:1995sp}%
  \BibitemOpen
  \bibfield  {author} {\bibinfo {author} {\bibfnamefont {R.}~\bibnamefont
  {Akhoury}}\ and\ \bibinfo {author} {\bibfnamefont {V.~I.}\ \bibnamefont
  {Zakharov}},\ }\bibfield  {title} {\bibinfo {title} {{On the universality of
  the leading, 1/Q power corrections in QCD}},\ }\href
  {https://doi.org/10.1016/0370-2693(95)00866-J} {\bibfield  {journal}
  {\bibinfo  {journal} {Phys. Lett. B}\ }\textbf {\bibinfo {volume} {357}},\
  \bibinfo {pages} {646} (\bibinfo {year} {1995})},\ \Eprint
  {https://arxiv.org/abs/hep-ph/9504248} {arXiv:hep-ph/9504248} \BibitemShut
  {NoStop}%
\bibitem [{\citenamefont {Nason}\ and\ \citenamefont
  {Seymour}(1995)}]{Nason:1995np}%
  \BibitemOpen
  \bibfield  {author} {\bibinfo {author} {\bibfnamefont {P.}~\bibnamefont
  {Nason}}\ and\ \bibinfo {author} {\bibfnamefont {M.~H.}\ \bibnamefont
  {Seymour}},\ }\bibfield  {title} {\bibinfo {title} {{Infrared renormalons and
  power suppressed effects in e+ e- jet events}},\ }\href
  {https://doi.org/10.1016/0550-3213(95)00461-Z} {\bibfield  {journal}
  {\bibinfo  {journal} {Nucl. Phys. B}\ }\textbf {\bibinfo {volume} {454}},\
  \bibinfo {pages} {291} (\bibinfo {year} {1995})},\ \Eprint
  {https://arxiv.org/abs/hep-ph/9506317} {arXiv:hep-ph/9506317} \BibitemShut
  {NoStop}%
\bibitem [{\citenamefont {Dokshitzer}\ \emph {et~al.}(1996)\citenamefont
  {Dokshitzer}, \citenamefont {Marchesini},\ and\ \citenamefont
  {Webber}}]{Dokshitzer:1995qm}%
  \BibitemOpen
  \bibfield  {author} {\bibinfo {author} {\bibfnamefont {Y.~L.}\ \bibnamefont
  {Dokshitzer}}, \bibinfo {author} {\bibfnamefont {G.}~\bibnamefont
  {Marchesini}},\ and\ \bibinfo {author} {\bibfnamefont {B.~R.}\ \bibnamefont
  {Webber}},\ }\bibfield  {title} {\bibinfo {title} {{Dispersive approach to
  power behaved contributions in QCD hard processes}},\ }\href
  {https://doi.org/10.1016/0550-3213(96)00155-1} {\bibfield  {journal}
  {\bibinfo  {journal} {Nucl. Phys. B}\ }\textbf {\bibinfo {volume} {469}},\
  \bibinfo {pages} {93} (\bibinfo {year} {1996})},\ \Eprint
  {https://arxiv.org/abs/hep-ph/9512336} {arXiv:hep-ph/9512336} \BibitemShut
  {NoStop}%
\bibitem [{\citenamefont {Beneke}\ \emph {et~al.}(1997)\citenamefont {Beneke},
  \citenamefont {Braun},\ and\ \citenamefont {Magnea}}]{Beneke:1997sr}%
  \BibitemOpen
  \bibfield  {author} {\bibinfo {author} {\bibfnamefont {M.}~\bibnamefont
  {Beneke}}, \bibinfo {author} {\bibfnamefont {V.~M.}\ \bibnamefont {Braun}},\
  and\ \bibinfo {author} {\bibfnamefont {L.}~\bibnamefont {Magnea}},\
  }\bibfield  {title} {\bibinfo {title} {{Phenomenology of power corrections in
  fragmentation processes in e+ e- annihilation}},\ }\href
  {https://doi.org/10.1016/S0550-3213(97)00251-4} {\bibfield  {journal}
  {\bibinfo  {journal} {Nucl. Phys. B}\ }\textbf {\bibinfo {volume} {497}},\
  \bibinfo {pages} {297} (\bibinfo {year} {1997})},\ \Eprint
  {https://arxiv.org/abs/hep-ph/9701309} {arXiv:hep-ph/9701309} \BibitemShut
  {NoStop}%
\bibitem [{\citenamefont {Dokshitzer}\ and\ \citenamefont
  {Webber}(1997)}]{Dokshitzer:1997ew}%
  \BibitemOpen
  \bibfield  {author} {\bibinfo {author} {\bibfnamefont {Y.~L.}\ \bibnamefont
  {Dokshitzer}}\ and\ \bibinfo {author} {\bibfnamefont {B.~R.}\ \bibnamefont
  {Webber}},\ }\bibfield  {title} {\bibinfo {title} {{Power corrections to
  event shape distributions}},\ }\href
  {https://doi.org/10.1016/S0370-2693(97)00573-X} {\bibfield  {journal}
  {\bibinfo  {journal} {Phys. Lett. B}\ }\textbf {\bibinfo {volume} {404}},\
  \bibinfo {pages} {321} (\bibinfo {year} {1997})},\ \Eprint
  {https://arxiv.org/abs/hep-ph/9704298} {arXiv:hep-ph/9704298} \BibitemShut
  {NoStop}%
\bibitem [{\citenamefont {Dokshitzer}\ \emph
  {et~al.}(1998{\natexlab{a}})\citenamefont {Dokshitzer}, \citenamefont
  {Lucenti}, \citenamefont {Marchesini},\ and\ \citenamefont
  {Salam}}]{Dokshitzer:1997iz}%
  \BibitemOpen
  \bibfield  {author} {\bibinfo {author} {\bibfnamefont {Y.~L.}\ \bibnamefont
  {Dokshitzer}}, \bibinfo {author} {\bibfnamefont {A.}~\bibnamefont {Lucenti}},
  \bibinfo {author} {\bibfnamefont {G.}~\bibnamefont {Marchesini}},\ and\
  \bibinfo {author} {\bibfnamefont {G.~P.}\ \bibnamefont {Salam}},\ }\bibfield
  {title} {\bibinfo {title} {{Universality of 1/Q corrections to jet-shape
  observables rescued}},\ }\href
  {https://doi.org/10.1016/S0550-3213(97)00650-0} {\bibfield  {journal}
  {\bibinfo  {journal} {Nucl. Phys. B}\ }\textbf {\bibinfo {volume} {511}},\
  \bibinfo {pages} {396} (\bibinfo {year} {1998}{\natexlab{a}})},\ \bibinfo
  {note} {[Erratum: Nucl.Phys.B 593, 729--730 (2001)]},\ \Eprint
  {https://arxiv.org/abs/hep-ph/9707532} {arXiv:hep-ph/9707532} \BibitemShut
  {NoStop}%
\bibitem [{\citenamefont {Dokshitzer}\ \emph
  {et~al.}(1998{\natexlab{b}})\citenamefont {Dokshitzer}, \citenamefont
  {Lucenti}, \citenamefont {Marchesini},\ and\ \citenamefont
  {Salam}}]{Dokshitzer:1998pt}%
  \BibitemOpen
  \bibfield  {author} {\bibinfo {author} {\bibfnamefont {Y.~L.}\ \bibnamefont
  {Dokshitzer}}, \bibinfo {author} {\bibfnamefont {A.}~\bibnamefont {Lucenti}},
  \bibinfo {author} {\bibfnamefont {G.}~\bibnamefont {Marchesini}},\ and\
  \bibinfo {author} {\bibfnamefont {G.~P.}\ \bibnamefont {Salam}},\ }\bibfield
  {title} {\bibinfo {title} {{On the universality of the Milan factor for 1 / Q
  power corrections to jet shapes}},\ }\href
  {https://doi.org/10.1088/1126-6708/1998/05/003} {\bibfield  {journal}
  {\bibinfo  {journal} {JHEP}\ }\textbf {\bibinfo {volume} {05}},\ \bibinfo
  {pages} {003}},\ \Eprint {https://arxiv.org/abs/hep-ph/9802381}
  {arXiv:hep-ph/9802381} \BibitemShut {NoStop}%
\bibitem [{\citenamefont {Dasgupta}\ and\ \citenamefont
  {Webber}(1998)}]{Dasgupta:1998xt}%
  \BibitemOpen
  \bibfield  {author} {\bibinfo {author} {\bibfnamefont {M.}~\bibnamefont
  {Dasgupta}}\ and\ \bibinfo {author} {\bibfnamefont {B.~R.}\ \bibnamefont
  {Webber}},\ }\bibfield  {title} {\bibinfo {title} {{Two loop enhancement
  factor for 1 / Q corrections to event shapes in deep inelastic scattering}},\
  }\href {https://doi.org/10.1088/1126-6708/1998/10/001} {\bibfield  {journal}
  {\bibinfo  {journal} {JHEP}\ }\textbf {\bibinfo {volume} {10}},\ \bibinfo
  {pages} {001}},\ \Eprint {https://arxiv.org/abs/hep-ph/9809247}
  {arXiv:hep-ph/9809247} \BibitemShut {NoStop}%
\bibitem [{\citenamefont {Dasgupta}\ \emph {et~al.}(1999)\citenamefont
  {Dasgupta}, \citenamefont {Magnea},\ and\ \citenamefont
  {Smye}}]{Dasgupta:1999mb}%
  \BibitemOpen
  \bibfield  {author} {\bibinfo {author} {\bibfnamefont {M.}~\bibnamefont
  {Dasgupta}}, \bibinfo {author} {\bibfnamefont {L.}~\bibnamefont {Magnea}},\
  and\ \bibinfo {author} {\bibfnamefont {G.}~\bibnamefont {Smye}},\ }\bibfield
  {title} {\bibinfo {title} {{Universality of 1/Q corrections revisited}},\
  }\href {https://doi.org/10.1088/1126-6708/1999/11/025} {\bibfield  {journal}
  {\bibinfo  {journal} {JHEP}\ }\textbf {\bibinfo {volume} {11}},\ \bibinfo
  {pages} {025}},\ \Eprint {https://arxiv.org/abs/hep-ph/9911316}
  {arXiv:hep-ph/9911316} \BibitemShut {NoStop}%
\bibitem [{\citenamefont {Korchemsky}\ and\ \citenamefont
  {Sterman}(1999)}]{Korchemsky:1999kt}%
  \BibitemOpen
  \bibfield  {author} {\bibinfo {author} {\bibfnamefont {G.~P.}\ \bibnamefont
  {Korchemsky}}\ and\ \bibinfo {author} {\bibfnamefont {G.~F.}\ \bibnamefont
  {Sterman}},\ }\bibfield  {title} {\bibinfo {title} {{Power corrections to
  event shapes and factorization}},\ }\href
  {https://doi.org/10.1016/S0550-3213(99)00308-9} {\bibfield  {journal}
  {\bibinfo  {journal} {Nucl. Phys. B}\ }\textbf {\bibinfo {volume} {555}},\
  \bibinfo {pages} {335} (\bibinfo {year} {1999})},\ \Eprint
  {https://arxiv.org/abs/hep-ph/9902341} {arXiv:hep-ph/9902341} \BibitemShut
  {NoStop}%
\bibitem [{\citenamefont {Dokshitzer}\ \emph {et~al.}(1999)\citenamefont
  {Dokshitzer}, \citenamefont {Marchesini},\ and\ \citenamefont
  {Webber}}]{Dokshitzer:1999sh}%
  \BibitemOpen
  \bibfield  {author} {\bibinfo {author} {\bibfnamefont {Y.~L.}\ \bibnamefont
  {Dokshitzer}}, \bibinfo {author} {\bibfnamefont {G.}~\bibnamefont
  {Marchesini}},\ and\ \bibinfo {author} {\bibfnamefont {B.~R.}\ \bibnamefont
  {Webber}},\ }\bibfield  {title} {\bibinfo {title} {{Nonperturbative effects
  in the energy energy correlation}},\ }\href
  {https://doi.org/10.1088/1126-6708/1999/07/012} {\bibfield  {journal}
  {\bibinfo  {journal} {JHEP}\ }\textbf {\bibinfo {volume} {07}},\ \bibinfo
  {pages} {012}},\ \Eprint {https://arxiv.org/abs/hep-ph/9905339}
  {arXiv:hep-ph/9905339} \BibitemShut {NoStop}%
\bibitem [{\citenamefont {Belitsky}\ \emph {et~al.}(2001)\citenamefont
  {Belitsky}, \citenamefont {Korchemsky},\ and\ \citenamefont
  {Sterman}}]{Belitsky:2001ij}%
  \BibitemOpen
  \bibfield  {author} {\bibinfo {author} {\bibfnamefont {A.~V.}\ \bibnamefont
  {Belitsky}}, \bibinfo {author} {\bibfnamefont {G.~P.}\ \bibnamefont
  {Korchemsky}},\ and\ \bibinfo {author} {\bibfnamefont {G.~F.}\ \bibnamefont
  {Sterman}},\ }\bibfield  {title} {\bibinfo {title} {{Energy flow in QCD and
  event shape functions}},\ }\href
  {https://doi.org/10.1016/S0370-2693(01)00899-1} {\bibfield  {journal}
  {\bibinfo  {journal} {Phys. Lett. B}\ }\textbf {\bibinfo {volume} {515}},\
  \bibinfo {pages} {297} (\bibinfo {year} {2001})},\ \Eprint
  {https://arxiv.org/abs/hep-ph/0106308} {arXiv:hep-ph/0106308} \BibitemShut
  {NoStop}%
\bibitem [{\citenamefont {Nason}\ and\ \citenamefont
  {Zanderighi}(2023)}]{Nason:2023asn}%
  \BibitemOpen
  \bibfield  {author} {\bibinfo {author} {\bibfnamefont {P.}~\bibnamefont
  {Nason}}\ and\ \bibinfo {author} {\bibfnamefont {G.}~\bibnamefont
  {Zanderighi}},\ }\bibfield  {title} {\bibinfo {title} {{Fits of
  {\ensuremath{\alpha}}$_{s}$ using power corrections in the three-jet
  region}},\ }\href {https://doi.org/10.1007/JHEP06(2023)058} {\bibfield
  {journal} {\bibinfo  {journal} {JHEP}\ }\textbf {\bibinfo {volume} {06}},\
  \bibinfo {pages} {058}},\ \Eprint {https://arxiv.org/abs/2301.03607}
  {arXiv:2301.03607 [hep-ph]} \BibitemShut {NoStop}%
\bibitem [{\citenamefont {Bell}\ \emph {et~al.}(2024)\citenamefont {Bell},
  \citenamefont {Lee}, \citenamefont {Makris}, \citenamefont {Talbert},\ and\
  \citenamefont {Yan}}]{Bell:2023dqs}%
  \BibitemOpen
  \bibfield  {author} {\bibinfo {author} {\bibfnamefont {G.}~\bibnamefont
  {Bell}}, \bibinfo {author} {\bibfnamefont {C.}~\bibnamefont {Lee}}, \bibinfo
  {author} {\bibfnamefont {Y.}~\bibnamefont {Makris}}, \bibinfo {author}
  {\bibfnamefont {J.}~\bibnamefont {Talbert}},\ and\ \bibinfo {author}
  {\bibfnamefont {B.}~\bibnamefont {Yan}},\ }\bibfield  {title} {\bibinfo
  {title} {{Effects of renormalon scheme and perturbative scale choices on
  determinations of the strong coupling from e+e- event shapes}},\ }\href
  {https://doi.org/10.1103/PhysRevD.109.094008} {\bibfield  {journal} {\bibinfo
   {journal} {Phys. Rev. D}\ }\textbf {\bibinfo {volume} {109}},\ \bibinfo
  {pages} {094008} (\bibinfo {year} {2024})},\ \Eprint
  {https://arxiv.org/abs/2311.03990} {arXiv:2311.03990 [hep-ph]} \BibitemShut
  {NoStop}%
\bibitem [{\citenamefont {Benitez}\ \emph
  {et~al.}(2025{\natexlab{a}})\citenamefont {Benitez}, \citenamefont {Hoang},
  \citenamefont {Mateu}, \citenamefont {Stewart},\ and\ \citenamefont
  {Vita}}]{Benitez:2024nav}%
  \BibitemOpen
  \bibfield  {author} {\bibinfo {author} {\bibfnamefont {M.~A.}\ \bibnamefont
  {Benitez}}, \bibinfo {author} {\bibfnamefont {A.~H.}\ \bibnamefont {Hoang}},
  \bibinfo {author} {\bibfnamefont {V.}~\bibnamefont {Mateu}}, \bibinfo
  {author} {\bibfnamefont {I.~W.}\ \bibnamefont {Stewart}},\ and\ \bibinfo
  {author} {\bibfnamefont {G.}~\bibnamefont {Vita}},\ }\bibfield  {title}
  {\bibinfo {title} {{On determining {\ensuremath{\alpha}}$_{s}$(m$_{Z}$) from
  dijets in e$^{+}$e$^{-}$ thrust}},\ }\href
  {https://doi.org/10.1007/JHEP07(2025)249} {\bibfield  {journal} {\bibinfo
  {journal} {JHEP}\ }\textbf {\bibinfo {volume} {07}},\ \bibinfo {pages}
  {249}},\ \Eprint {https://arxiv.org/abs/2412.15164} {arXiv:2412.15164
  [hep-ph]} \BibitemShut {NoStop}%
\bibitem [{\citenamefont {Nason}\ and\ \citenamefont
  {Zanderighi}(2025)}]{Nason:2025qbx}%
  \BibitemOpen
  \bibfield  {author} {\bibinfo {author} {\bibfnamefont {P.}~\bibnamefont
  {Nason}}\ and\ \bibinfo {author} {\bibfnamefont {G.}~\bibnamefont
  {Zanderighi}},\ }\bibfield  {title} {\bibinfo {title} {{Fits of
  {\ensuremath{\alpha}}$_{s}$ from event-shapes in the three-jet region:
  extension to all energies}},\ }\href
  {https://doi.org/10.1007/JHEP06(2025)200} {\bibfield  {journal} {\bibinfo
  {journal} {JHEP}\ }\textbf {\bibinfo {volume} {06}},\ \bibinfo {pages}
  {200}},\ \Eprint {https://arxiv.org/abs/2501.18173} {arXiv:2501.18173
  [hep-ph]} \BibitemShut {NoStop}%
\bibitem [{\citenamefont {Aglietti}\ \emph {et~al.}(2025)\citenamefont
  {Aglietti}, \citenamefont {Ferrera}, \citenamefont {Ju},\ and\ \citenamefont
  {Miao}}]{Aglietti:2025jdj}%
  \BibitemOpen
  \bibfield  {author} {\bibinfo {author} {\bibfnamefont {U.~G.}\ \bibnamefont
  {Aglietti}}, \bibinfo {author} {\bibfnamefont {G.}~\bibnamefont {Ferrera}},
  \bibinfo {author} {\bibfnamefont {W.-L.}\ \bibnamefont {Ju}},\ and\ \bibinfo
  {author} {\bibfnamefont {J.}~\bibnamefont {Miao}},\ }\bibfield  {title}
  {\bibinfo {title} {{Thrust Distribution in Electron-Positron Annihilation at
  Full Next-to-Next-to-Next-to-Leading-Logarithmic Accuracy Including
  Next-to-Next-to-Leading-Order Terms in QCD}},\ }\href
  {https://doi.org/10.1103/dv7n-qvyp} {\bibfield  {journal} {\bibinfo
  {journal} {Phys. Rev. Lett.}\ }\textbf {\bibinfo {volume} {134}},\ \bibinfo
  {pages} {251904} (\bibinfo {year} {2025})},\ \Eprint
  {https://arxiv.org/abs/2502.01570} {arXiv:2502.01570 [hep-ph]} \BibitemShut
  {NoStop}%
\bibitem [{\citenamefont {Benitez}\ \emph
  {et~al.}(2025{\natexlab{b}})\citenamefont {Benitez}, \citenamefont
  {Bhattacharya}, \citenamefont {Hoang}, \citenamefont {Mateu}, \citenamefont
  {Schwartz}, \citenamefont {Stewart},\ and\ \citenamefont
  {Zhang}}]{Benitez:2025vsp}%
  \BibitemOpen
  \bibfield  {author} {\bibinfo {author} {\bibfnamefont {M.~A.}\ \bibnamefont
  {Benitez}}, \bibinfo {author} {\bibfnamefont {A.}~\bibnamefont
  {Bhattacharya}}, \bibinfo {author} {\bibfnamefont {A.~H.}\ \bibnamefont
  {Hoang}}, \bibinfo {author} {\bibfnamefont {V.}~\bibnamefont {Mateu}},
  \bibinfo {author} {\bibfnamefont {M.~D.}\ \bibnamefont {Schwartz}}, \bibinfo
  {author} {\bibfnamefont {I.~W.}\ \bibnamefont {Stewart}},\ and\ \bibinfo
  {author} {\bibfnamefont {X.}~\bibnamefont {Zhang}},\ }\bibfield  {title}
  {\bibinfo {title} {{A Precise Determination of $\alpha_s$ from the Heavy Jet
  Mass Distribution}},\ }\href@noop {} {\  (\bibinfo {year}
  {2025}{\natexlab{b}})},\ \Eprint {https://arxiv.org/abs/2502.12253}
  {arXiv:2502.12253 [hep-ph]} \BibitemShut {NoStop}%
\bibitem [{\citenamefont {Navas}\ \emph {et~al.}(2024)\citenamefont {Navas}
  \emph {et~al.}}]{ParticleDataGroup:2024cfk}%
  \BibitemOpen
  \bibfield  {author} {\bibinfo {author} {\bibfnamefont {S.}~\bibnamefont
  {Navas}} \emph {et~al.} (\bibinfo {collaboration} {Particle Data Group}),\
  }\bibfield  {title} {\bibinfo {title} {{Review of particle physics}},\ }\href
  {https://doi.org/10.1103/PhysRevD.110.030001} {\bibfield  {journal} {\bibinfo
   {journal} {Phys. Rev. D}\ }\textbf {\bibinfo {volume} {110}},\ \bibinfo
  {pages} {030001} (\bibinfo {year} {2024})}\BibitemShut {NoStop}%
\bibitem [{\citenamefont {Mateu}\ \emph {et~al.}(2013)\citenamefont {Mateu},
  \citenamefont {Stewart},\ and\ \citenamefont {Thaler}}]{Mateu:2012nk}%
  \BibitemOpen
  \bibfield  {author} {\bibinfo {author} {\bibfnamefont {V.}~\bibnamefont
  {Mateu}}, \bibinfo {author} {\bibfnamefont {I.~W.}\ \bibnamefont {Stewart}},\
  and\ \bibinfo {author} {\bibfnamefont {J.}~\bibnamefont {Thaler}},\
  }\bibfield  {title} {\bibinfo {title} {{Power Corrections to Event Shapes
  with Mass-Dependent Operators}},\ }\href
  {https://doi.org/10.1103/PhysRevD.87.014025} {\bibfield  {journal} {\bibinfo
  {journal} {Phys. Rev. D}\ }\textbf {\bibinfo {volume} {87}},\ \bibinfo
  {pages} {014025} (\bibinfo {year} {2013})},\ \Eprint
  {https://arxiv.org/abs/1209.3781} {arXiv:1209.3781 [hep-ph]} \BibitemShut
  {NoStop}%
\bibitem [{\citenamefont {Chen}\ \emph {et~al.}(2024)\citenamefont {Chen},
  \citenamefont {Monni}, \citenamefont {Xu},\ and\ \citenamefont
  {Zhu}}]{Chen:2024nyc}%
  \BibitemOpen
  \bibfield  {author} {\bibinfo {author} {\bibfnamefont {H.}~\bibnamefont
  {Chen}}, \bibinfo {author} {\bibfnamefont {P.~F.}\ \bibnamefont {Monni}},
  \bibinfo {author} {\bibfnamefont {Z.}~\bibnamefont {Xu}},\ and\ \bibinfo
  {author} {\bibfnamefont {H.~X.}\ \bibnamefont {Zhu}},\ }\bibfield  {title}
  {\bibinfo {title} {{Scaling Violation in Power Corrections to Energy
  Correlators from the Light-Ray Operator Product Expansion}},\ }\href
  {https://doi.org/10.1103/PhysRevLett.133.231901} {\bibfield  {journal}
  {\bibinfo  {journal} {Phys. Rev. Lett.}\ }\textbf {\bibinfo {volume} {133}},\
  \bibinfo {pages} {231901} (\bibinfo {year} {2024})},\ \Eprint
  {https://arxiv.org/abs/2406.06668} {arXiv:2406.06668 [hep-ph]} \BibitemShut
  {NoStop}%
\bibitem [{\citenamefont {Chang}\ \emph
  {et~al.}(2026{\natexlab{a}})\citenamefont {Chang}, \citenamefont {Chen},
  \citenamefont {Liu}, \citenamefont {Simmons-Duffin}, \citenamefont {Yuan},\
  and\ \citenamefont {Zhu}}]{Chang:2025kgq}%
  \BibitemOpen
  \bibfield  {author} {\bibinfo {author} {\bibfnamefont {C.-H.}\ \bibnamefont
  {Chang}}, \bibinfo {author} {\bibfnamefont {H.}~\bibnamefont {Chen}},
  \bibinfo {author} {\bibfnamefont {X.}~\bibnamefont {Liu}}, \bibinfo {author}
  {\bibfnamefont {D.}~\bibnamefont {Simmons-Duffin}}, \bibinfo {author}
  {\bibfnamefont {F.}~\bibnamefont {Yuan}},\ and\ \bibinfo {author}
  {\bibfnamefont {H.~X.}\ \bibnamefont {Zhu}},\ }\bibfield  {title} {\bibinfo
  {title} {{Quantum Scaling in Energy Correlators beyond the Confinement
  Transition}},\ }\href {https://doi.org/10.1103/9ml8-xkfc} {\bibfield
  {journal} {\bibinfo  {journal} {Phys. Rev. Lett.}\ }\textbf {\bibinfo
  {volume} {136}},\ \bibinfo {pages} {081903} (\bibinfo {year}
  {2026}{\natexlab{a}})},\ \Eprint {https://arxiv.org/abs/2507.15923}
  {arXiv:2507.15923 [hep-ph]} \BibitemShut {NoStop}%
\bibitem [{\citenamefont {Lee}\ \emph {et~al.}(2024)\citenamefont {Lee},
  \citenamefont {Pathak}, \citenamefont {Stewart},\ and\ \citenamefont
  {Sun}}]{Lee:2024esz}%
  \BibitemOpen
  \bibfield  {author} {\bibinfo {author} {\bibfnamefont {K.}~\bibnamefont
  {Lee}}, \bibinfo {author} {\bibfnamefont {A.}~\bibnamefont {Pathak}},
  \bibinfo {author} {\bibfnamefont {I.~W.}\ \bibnamefont {Stewart}},\ and\
  \bibinfo {author} {\bibfnamefont {Z.}~\bibnamefont {Sun}},\ }\bibfield
  {title} {\bibinfo {title} {{Nonperturbative Effects in Energy Correlators:
  From Characterizing Confinement Transition to Improving
  {\ensuremath{\alpha}}s Extraction}},\ }\href
  {https://doi.org/10.1103/PhysRevLett.133.231902} {\bibfield  {journal}
  {\bibinfo  {journal} {Phys. Rev. Lett.}\ }\textbf {\bibinfo {volume} {133}},\
  \bibinfo {pages} {231902} (\bibinfo {year} {2024})},\ \Eprint
  {https://arxiv.org/abs/2405.19396} {arXiv:2405.19396 [hep-ph]} \BibitemShut
  {NoStop}%
\bibitem [{\citenamefont {Lee}\ and\ \citenamefont
  {Stewart}(2026)}]{Lee:2025okn}%
  \BibitemOpen
  \bibfield  {author} {\bibinfo {author} {\bibfnamefont {K.}~\bibnamefont
  {Lee}}\ and\ \bibinfo {author} {\bibfnamefont {I.~W.}\ \bibnamefont
  {Stewart}},\ }\bibfield  {title} {\bibinfo {title} {{Dihadron Fragmentation
  and the Confinement Transition in Energy Correlators}},\ }\href
  {https://doi.org/10.1103/m18j-xypt} {\bibfield  {journal} {\bibinfo
  {journal} {Phys. Rev. Lett.}\ }\textbf {\bibinfo {volume} {136}},\ \bibinfo
  {pages} {081902} (\bibinfo {year} {2026})},\ \Eprint
  {https://arxiv.org/abs/2507.11495} {arXiv:2507.11495 [hep-ph]} \BibitemShut
  {NoStop}%
\bibitem [{\citenamefont {Guo}\ \emph {et~al.}(2026)\citenamefont {Guo},
  \citenamefont {Yuan},\ and\ \citenamefont {Zhao}}]{Guo:2025zwb}%
  \BibitemOpen
  \bibfield  {author} {\bibinfo {author} {\bibfnamefont {Y.}~\bibnamefont
  {Guo}}, \bibinfo {author} {\bibfnamefont {F.}~\bibnamefont {Yuan}},\ and\
  \bibinfo {author} {\bibfnamefont {W.}~\bibnamefont {Zhao}},\ }\bibfield
  {title} {\bibinfo {title} {{Factorization and Resummation for the Nearside
  Energy-Energy Correlators}},\ }\href {https://doi.org/10.1103/4qkq-x5st}
  {\bibfield  {journal} {\bibinfo  {journal} {Phys. Rev. Lett.}\ }\textbf
  {\bibinfo {volume} {136}},\ \bibinfo {pages} {081904} (\bibinfo {year}
  {2026})},\ \Eprint {https://arxiv.org/abs/2507.15820} {arXiv:2507.15820
  [hep-ph]} \BibitemShut {NoStop}%
\bibitem [{\citenamefont {Kang}\ \emph {et~al.}(2026)\citenamefont {Kang},
  \citenamefont {Metz}, \citenamefont {Pitonyak},\ and\ \citenamefont
  {Zhang}}]{Kang:2025zto}%
  \BibitemOpen
  \bibfield  {author} {\bibinfo {author} {\bibfnamefont {Z.-B.}\ \bibnamefont
  {Kang}}, \bibinfo {author} {\bibfnamefont {A.}~\bibnamefont {Metz}}, \bibinfo
  {author} {\bibfnamefont {D.}~\bibnamefont {Pitonyak}},\ and\ \bibinfo
  {author} {\bibfnamefont {C.}~\bibnamefont {Zhang}},\ }\bibfield  {title}
  {\bibinfo {title} {{Dihadron Fragmentation Framework for Near-Side
  Energy-Energy Correlators}},\ }\href {https://doi.org/10.1103/jnl4-x77t}
  {\bibfield  {journal} {\bibinfo  {journal} {Phys. Rev. Lett.}\ }\textbf
  {\bibinfo {volume} {136}},\ \bibinfo {pages} {081905} (\bibinfo {year}
  {2026})},\ \Eprint {https://arxiv.org/abs/2507.17444} {arXiv:2507.17444
  [hep-ph]} \BibitemShut {NoStop}%
\bibitem [{\citenamefont {Herrmann}\ \emph {et~al.}(2025)\citenamefont
  {Herrmann}, \citenamefont {Kang}, \citenamefont {Penttala},\ and\
  \citenamefont {Zhang}}]{Herrmann:2025fqy}%
  \BibitemOpen
  \bibfield  {author} {\bibinfo {author} {\bibfnamefont {E.}~\bibnamefont
  {Herrmann}}, \bibinfo {author} {\bibfnamefont {Z.-B.}\ \bibnamefont {Kang}},
  \bibinfo {author} {\bibfnamefont {J.}~\bibnamefont {Penttala}},\ and\
  \bibinfo {author} {\bibfnamefont {C.}~\bibnamefont {Zhang}},\ }\bibfield
  {title} {\bibinfo {title} {{Collinear limit of the energy-energy correlator
  in $e^+ e^-$ collisions: transition from perturbative to non-perturbative
  regimes}},\ }\href@noop {} {\  (\bibinfo {year} {2025})},\ \Eprint
  {https://arxiv.org/abs/2507.17704} {arXiv:2507.17704 [hep-ph]} \BibitemShut
  {NoStop}%
\bibitem [{\citenamefont {Farren-Colloty}\ \emph {et~al.}(2025)\citenamefont
  {Farren-Colloty}, \citenamefont {Helliwell}, \citenamefont {Patel},
  \citenamefont {Salam},\ and\ \citenamefont
  {Zanoli}}]{Farren-Colloty:2025amh}%
  \BibitemOpen
  \bibfield  {author} {\bibinfo {author} {\bibfnamefont {C.}~\bibnamefont
  {Farren-Colloty}}, \bibinfo {author} {\bibfnamefont {J.}~\bibnamefont
  {Helliwell}}, \bibinfo {author} {\bibfnamefont {R.}~\bibnamefont {Patel}},
  \bibinfo {author} {\bibfnamefont {G.~P.}\ \bibnamefont {Salam}},\ and\
  \bibinfo {author} {\bibfnamefont {S.}~\bibnamefont {Zanoli}},\ }\bibfield
  {title} {\bibinfo {title} {{Anomalous scaling of linear power corrections}},\
  }\href@noop {} {\  (\bibinfo {year} {2025})},\ \Eprint
  {https://arxiv.org/abs/2507.18696} {arXiv:2507.18696 [hep-ph]} \BibitemShut
  {NoStop}%
\bibitem [{\citenamefont {Dasgupta}\ and\ \citenamefont
  {Hounat}(2025)}]{Dasgupta:2024znl}%
  \BibitemOpen
  \bibfield  {author} {\bibinfo {author} {\bibfnamefont {M.}~\bibnamefont
  {Dasgupta}}\ and\ \bibinfo {author} {\bibfnamefont {F.}~\bibnamefont
  {Hounat}},\ }\bibfield  {title} {\bibinfo {title} {{Exploring soft anomalous
  dimensions for 1/Q power corrections}},\ }\href
  {https://doi.org/10.1007/JHEP09(2025)060} {\bibfield  {journal} {\bibinfo
  {journal} {JHEP}\ }\textbf {\bibinfo {volume} {09}},\ \bibinfo {pages}
  {060}},\ \Eprint {https://arxiv.org/abs/2411.16867} {arXiv:2411.16867
  [hep-ph]} \BibitemShut {NoStop}%
\bibitem [{\citenamefont {Banfi}\ and\ \citenamefont
  {El-Menoufi}(2025)}]{Banfi:2025crj}%
  \BibitemOpen
  \bibfield  {author} {\bibinfo {author} {\bibfnamefont {A.}~\bibnamefont
  {Banfi}}\ and\ \bibinfo {author} {\bibfnamefont {B.~K.}\ \bibnamefont
  {El-Menoufi}},\ }\bibfield  {title} {\bibinfo {title} {{Characterising
  hadronisation across the phase space}},\ }\href@noop {} {\  (\bibinfo {year}
  {2025})},\ \Eprint {https://arxiv.org/abs/2511.17708} {arXiv:2511.17708
  [hep-ph]} \BibitemShut {NoStop}%
\bibitem [{\citenamefont {Kuraev}\ \emph {et~al.}(1977)\citenamefont {Kuraev},
  \citenamefont {Lipatov},\ and\ \citenamefont {Fadin}}]{Kuraev:1977fs}%
  \BibitemOpen
  \bibfield  {author} {\bibinfo {author} {\bibfnamefont {E.~A.}\ \bibnamefont
  {Kuraev}}, \bibinfo {author} {\bibfnamefont {L.~N.}\ \bibnamefont
  {Lipatov}},\ and\ \bibinfo {author} {\bibfnamefont {V.~S.}\ \bibnamefont
  {Fadin}},\ }\bibfield  {title} {\bibinfo {title} {{The Pomeranchuk
  singularity in nonabelian gauge theories}},\ }\href@noop {} {\bibfield
  {journal} {\bibinfo  {journal} {Sov. Phys. JETP}\ }\textbf {\bibinfo {volume}
  {45}},\ \bibinfo {pages} {199} (\bibinfo {year} {1977})}\BibitemShut
  {NoStop}%
\bibitem [{\citenamefont {Balitsky}\ and\ \citenamefont
  {Lipatov}(1978)}]{Balitsky:1978ic}%
  \BibitemOpen
  \bibfield  {author} {\bibinfo {author} {\bibfnamefont {I.~I.}\ \bibnamefont
  {Balitsky}}\ and\ \bibinfo {author} {\bibfnamefont {L.~N.}\ \bibnamefont
  {Lipatov}},\ }\bibfield  {title} {\bibinfo {title} {{The Pomeranchuk
  Singularity in Quantum Chromodynamics}},\ }\href@noop {} {\bibfield
  {journal} {\bibinfo  {journal} {Sov. J. Nucl. Phys.}\ }\textbf {\bibinfo
  {volume} {28}},\ \bibinfo {pages} {822} (\bibinfo {year} {1978})}\BibitemShut
  {NoStop}%
\bibitem [{\citenamefont {Lipatov}(1997)}]{Lipatov:1996ts}%
  \BibitemOpen
  \bibfield  {author} {\bibinfo {author} {\bibfnamefont {L.~N.}\ \bibnamefont
  {Lipatov}},\ }\bibfield  {title} {\bibinfo {title} {{Small x physics in
  perturbative QCD}},\ }\href {https://doi.org/10.1016/S0370-1573(96)00045-2}
  {\bibfield  {journal} {\bibinfo  {journal} {Phys. Rept.}\ }\textbf {\bibinfo
  {volume} {286}},\ \bibinfo {pages} {131} (\bibinfo {year} {1997})},\ \Eprint
  {https://arxiv.org/abs/hep-ph/9610276} {arXiv:hep-ph/9610276} \BibitemShut
  {NoStop}%
\bibitem [{\citenamefont {Chang}\ \emph
  {et~al.}(2026{\natexlab{b}})\citenamefont {Chang}, \citenamefont {Chen},
  \citenamefont {Simmons-Duffin},\ and\ \citenamefont {Zhu}}]{Chang:2025zib}%
  \BibitemOpen
  \bibfield  {author} {\bibinfo {author} {\bibfnamefont {C.-H.}\ \bibnamefont
  {Chang}}, \bibinfo {author} {\bibfnamefont {H.}~\bibnamefont {Chen}},
  \bibinfo {author} {\bibfnamefont {D.}~\bibnamefont {Simmons-Duffin}},\ and\
  \bibinfo {author} {\bibfnamefont {H.~X.}\ \bibnamefont {Zhu}},\ }\bibfield
  {title} {\bibinfo {title} {{Seeing through the confinement screen: DGLAP/BFKL
  mixing and light-ray matching in QCD}},\ }\href
  {https://doi.org/10.1007/JHEP02(2026)251} {\bibfield  {journal} {\bibinfo
  {journal} {JHEP}\ }\textbf {\bibinfo {volume} {02}},\ \bibinfo {pages}
  {251}},\ \Eprint {https://arxiv.org/abs/2506.06431} {arXiv:2506.06431
  [hep-th]} \BibitemShut {NoStop}%
\bibitem [{\citenamefont {Schindler}\ \emph {et~al.}(2023)\citenamefont
  {Schindler}, \citenamefont {Stewart},\ and\ \citenamefont
  {Sun}}]{Schindler:2023cww}%
  \BibitemOpen
  \bibfield  {author} {\bibinfo {author} {\bibfnamefont {S.~T.}\ \bibnamefont
  {Schindler}}, \bibinfo {author} {\bibfnamefont {I.~W.}\ \bibnamefont
  {Stewart}},\ and\ \bibinfo {author} {\bibfnamefont {Z.}~\bibnamefont {Sun}},\
  }\bibfield  {title} {\bibinfo {title} {{Renormalons in the energy-energy
  correlator}},\ }\href {https://doi.org/10.1007/JHEP10(2023)187} {\bibfield
  {journal} {\bibinfo  {journal} {JHEP}\ }\textbf {\bibinfo {volume} {10}},\
  \bibinfo {pages} {187}},\ \bibinfo {note} {[Erratum: JHEP 10, 175 (2024)]},\
  \Eprint {https://arxiv.org/abs/2305.19311} {arXiv:2305.19311 [hep-ph]}
  \BibitemShut {NoStop}%
\bibitem [{\citenamefont {Lee}\ and\ \citenamefont
  {Sterman}(2007)}]{Lee:2006nr}%
  \BibitemOpen
  \bibfield  {author} {\bibinfo {author} {\bibfnamefont {C.}~\bibnamefont
  {Lee}}\ and\ \bibinfo {author} {\bibfnamefont {G.~F.}\ \bibnamefont
  {Sterman}},\ }\bibfield  {title} {\bibinfo {title} {{Momentum Flow
  Correlations from Event Shapes: Factorized Soft Gluons and Soft-Collinear
  Effective Theory}},\ }\href {https://doi.org/10.1103/PhysRevD.75.014022}
  {\bibfield  {journal} {\bibinfo  {journal} {Phys. Rev. D}\ }\textbf {\bibinfo
  {volume} {75}},\ \bibinfo {pages} {014022} (\bibinfo {year} {2007})},\
  \Eprint {https://arxiv.org/abs/hep-ph/0611061} {arXiv:hep-ph/0611061}
  \BibitemShut {NoStop}%
\bibitem [{\citenamefont {Salam}\ and\ \citenamefont
  {Wicke}(2001)}]{Salam:2001bd}%
  \BibitemOpen
  \bibfield  {author} {\bibinfo {author} {\bibfnamefont {G.~P.}\ \bibnamefont
  {Salam}}\ and\ \bibinfo {author} {\bibfnamefont {D.}~\bibnamefont {Wicke}},\
  }\bibfield  {title} {\bibinfo {title} {{Hadron masses and power corrections
  to event shapes}},\ }\href {https://doi.org/10.1088/1126-6708/2001/05/061}
  {\bibfield  {journal} {\bibinfo  {journal} {JHEP}\ }\textbf {\bibinfo
  {volume} {05}},\ \bibinfo {pages} {061}},\ \Eprint
  {https://arxiv.org/abs/hep-ph/0102343} {arXiv:hep-ph/0102343} \BibitemShut
  {NoStop}%
\bibitem [{Note1()}]{Note1}%
  \BibitemOpen
  \bibinfo {note} {In this Letter, we neglect the contribution from multi-leg
  Wilson line contribution.}\BibitemShut {Stop}%
\bibitem [{\citenamefont {Caron-Huot}\ \emph {et~al.}(2023)\citenamefont
  {Caron-Huot}, \citenamefont {Kologlu}, \citenamefont {Kravchuk},
  \citenamefont {Meltzer},\ and\ \citenamefont
  {Simmons-Duffin}}]{Caron-Huot:2022eqs}%
  \BibitemOpen
  \bibfield  {author} {\bibinfo {author} {\bibfnamefont {S.}~\bibnamefont
  {Caron-Huot}}, \bibinfo {author} {\bibfnamefont {M.}~\bibnamefont {Kologlu}},
  \bibinfo {author} {\bibfnamefont {P.}~\bibnamefont {Kravchuk}}, \bibinfo
  {author} {\bibfnamefont {D.}~\bibnamefont {Meltzer}},\ and\ \bibinfo {author}
  {\bibfnamefont {D.}~\bibnamefont {Simmons-Duffin}},\ }\bibfield  {title}
  {\bibinfo {title} {{Detectors in weakly-coupled field theories}},\ }\href
  {https://doi.org/10.1007/JHEP04(2023)014} {\bibfield  {journal} {\bibinfo
  {journal} {JHEP}\ }\textbf {\bibinfo {volume} {04}},\ \bibinfo {pages}
  {014}},\ \Eprint {https://arxiv.org/abs/2209.00008} {arXiv:2209.00008
  [hep-th]} \BibitemShut {NoStop}%
\bibitem [{\citenamefont {Giele}\ and\ \citenamefont
  {Glover}(1992)}]{Giele:1991vf}%
  \BibitemOpen
  \bibfield  {author} {\bibinfo {author} {\bibfnamefont {W.~T.}\ \bibnamefont
  {Giele}}\ and\ \bibinfo {author} {\bibfnamefont {E.~W.~N.}\ \bibnamefont
  {Glover}},\ }\bibfield  {title} {\bibinfo {title} {{Higher Order Corrections
  to Jet Cross Sections in e+ e- Annihilation}},\ }\href
  {https://doi.org/10.1103/PhysRevD.46.1980} {\bibfield  {journal} {\bibinfo
  {journal} {Phys. Rev. D}\ }\textbf {\bibinfo {volume} {46}},\ \bibinfo
  {pages} {1980} (\bibinfo {year} {1992})}\BibitemShut {NoStop}%
\bibitem [{\citenamefont {Kunszt}\ \emph {et~al.}(1994)\citenamefont {Kunszt},
  \citenamefont {Signer},\ and\ \citenamefont {Trocsanyi}}]{Kunszt:1994np}%
  \BibitemOpen
  \bibfield  {author} {\bibinfo {author} {\bibfnamefont {Z.}~\bibnamefont
  {Kunszt}}, \bibinfo {author} {\bibfnamefont {A.}~\bibnamefont {Signer}},\
  and\ \bibinfo {author} {\bibfnamefont {Z.}~\bibnamefont {Trocsanyi}},\
  }\bibfield  {title} {\bibinfo {title} {{Singular terms of helicity amplitudes
  at one loop in QCD and the soft limit of the cross-sections of multiparton
  processes}},\ }\href {https://doi.org/10.1016/0550-3213(94)90077-9}
  {\bibfield  {journal} {\bibinfo  {journal} {Nucl. Phys. B}\ }\textbf
  {\bibinfo {volume} {420}},\ \bibinfo {pages} {550} (\bibinfo {year}
  {1994})},\ \Eprint {https://arxiv.org/abs/hep-ph/9401294}
  {arXiv:hep-ph/9401294} \BibitemShut {NoStop}%
\bibitem [{\citenamefont {Catani}\ and\ \citenamefont
  {Seymour}(1997)}]{Catani:1996vz}%
  \BibitemOpen
  \bibfield  {author} {\bibinfo {author} {\bibfnamefont {S.}~\bibnamefont
  {Catani}}\ and\ \bibinfo {author} {\bibfnamefont {M.~H.}\ \bibnamefont
  {Seymour}},\ }\bibfield  {title} {\bibinfo {title} {{A General algorithm for
  calculating jet cross-sections in NLO QCD}},\ }\href
  {https://doi.org/10.1016/S0550-3213(96)00589-5} {\bibfield  {journal}
  {\bibinfo  {journal} {Nucl. Phys. B}\ }\textbf {\bibinfo {volume} {485}},\
  \bibinfo {pages} {291} (\bibinfo {year} {1997})},\ \bibinfo {note} {[Erratum:
  Nucl.Phys.B 510, 503--504 (1998)]},\ \Eprint
  {https://arxiv.org/abs/hep-ph/9605323} {arXiv:hep-ph/9605323} \BibitemShut
  {NoStop}%
\bibitem [{\citenamefont {Becher}\ and\ \citenamefont
  {Neubert}(2006)}]{Becher:2006nr}%
  \BibitemOpen
  \bibfield  {author} {\bibinfo {author} {\bibfnamefont {T.}~\bibnamefont
  {Becher}}\ and\ \bibinfo {author} {\bibfnamefont {M.}~\bibnamefont
  {Neubert}},\ }\bibfield  {title} {\bibinfo {title} {{Threshold resummation in
  momentum space from effective field theory}},\ }\href
  {https://doi.org/10.1103/PhysRevLett.97.082001} {\bibfield  {journal}
  {\bibinfo  {journal} {Phys. Rev. Lett.}\ }\textbf {\bibinfo {volume} {97}},\
  \bibinfo {pages} {082001} (\bibinfo {year} {2006})},\ \Eprint
  {https://arxiv.org/abs/hep-ph/0605050} {arXiv:hep-ph/0605050} \BibitemShut
  {NoStop}%
\bibitem [{\citenamefont {Becher}\ and\ \citenamefont
  {Neubert}(2009)}]{Becher:2009cu}%
  \BibitemOpen
  \bibfield  {author} {\bibinfo {author} {\bibfnamefont {T.}~\bibnamefont
  {Becher}}\ and\ \bibinfo {author} {\bibfnamefont {M.}~\bibnamefont
  {Neubert}},\ }\bibfield  {title} {\bibinfo {title} {{Infrared singularities
  of scattering amplitudes in perturbative QCD}},\ }\href
  {https://doi.org/10.1103/PhysRevLett.102.162001} {\bibfield  {journal}
  {\bibinfo  {journal} {Phys. Rev. Lett.}\ }\textbf {\bibinfo {volume} {102}},\
  \bibinfo {pages} {162001} (\bibinfo {year} {2009})},\ \bibinfo {note}
  {[Erratum: Phys.Rev.Lett. 111, 199905 (2013)]},\ \Eprint
  {https://arxiv.org/abs/0901.0722} {arXiv:0901.0722 [hep-ph]} \BibitemShut
  {NoStop}%
\bibitem [{\citenamefont {Bierlich}\ \emph {et~al.}(2022)\citenamefont
  {Bierlich} \emph {et~al.}}]{Bierlich:2022pfr}%
  \BibitemOpen
  \bibfield  {author} {\bibinfo {author} {\bibfnamefont {C.}~\bibnamefont
  {Bierlich}} \emph {et~al.},\ }\bibfield  {title} {\bibinfo {title} {{A
  comprehensive guide to the physics and usage of PYTHIA 8.3}},\ }\href
  {https://doi.org/10.21468/SciPostPhysCodeb.8} {\bibfield  {journal} {\bibinfo
   {journal} {SciPost Phys. Codeb.}\ }\textbf {\bibinfo {volume} {2022}},\
  \bibinfo {pages} {8} (\bibinfo {year} {2022})},\ \Eprint
  {https://arxiv.org/abs/2203.11601} {arXiv:2203.11601 [hep-ph]} \BibitemShut
  {NoStop}%
\bibitem [{\citenamefont {Kologlu}\ \emph {et~al.}(2020)\citenamefont
  {Kologlu}, \citenamefont {Kravchuk}, \citenamefont {Simmons-Duffin},\ and\
  \citenamefont {Zhiboedov}}]{Kologlu:2019bco}%
  \BibitemOpen
  \bibfield  {author} {\bibinfo {author} {\bibfnamefont {M.}~\bibnamefont
  {Kologlu}}, \bibinfo {author} {\bibfnamefont {P.}~\bibnamefont {Kravchuk}},
  \bibinfo {author} {\bibfnamefont {D.}~\bibnamefont {Simmons-Duffin}},\ and\
  \bibinfo {author} {\bibfnamefont {A.}~\bibnamefont {Zhiboedov}},\ }\bibfield
  {title} {\bibinfo {title} {{Shocks, Superconvergence, and a Stringy
  Equivalence Principle}},\ }\href {https://doi.org/10.1007/JHEP11(2020)096}
  {\bibfield  {journal} {\bibinfo  {journal} {JHEP}\ }\textbf {\bibinfo
  {volume} {11}},\ \bibinfo {pages} {096}},\ \Eprint
  {https://arxiv.org/abs/1904.05905} {arXiv:1904.05905 [hep-th]} \BibitemShut
  {NoStop}%
\end{thebibliography}%

\end{document}